\documentclass[apj]{emulateapj}

\shorttitle{Parallax distances to two binary MSPs}

\newcommand{\psrpi}{\ensuremath{\mathrm{PSR}\Large\pi}}
\newcommand{\mspsrpi}{\ensuremath{\mathrm{MSPSR}\Large\pi}}
\newcommand{\Msun}{\ensuremath{M_{\odot}}}
\newcommand{\psrone}{PSR J1022+1001}
\newcommand{\psrtwo}{PSR J2145--0750}

\newcommand{\degrees}{\ensuremath{^\circ}}

\newcommand{\til}{\raisebox{-0.7ex}{\ensuremath{\tilde{\;}}}}
\newcommand{\hst}{\textit{HST}}
\newcommand{\Mp}{\ensuremath{M_{\rm psr}}}
\newcommand{\Mc}{\ensuremath{M_c}}
\newcommand{\Rc}{\ensuremath{R_c}}
\newcommand{\Teff}{\ensuremath{T_{\rm eff}}}

\begin{document}

\title{Microarcsecond VLBI pulsar astrometry with \psrpi\ I. Two binary millisecond pulsars with white dwarf companions}

\author{
A. T. Deller\altaffilmark{1},
S. J. Vigeland\altaffilmark{2},
D. L. Kaplan\altaffilmark{2},
W. M. Goss\altaffilmark{3},
W. F. Brisken\altaffilmark{3},
S.~Chatterjee\altaffilmark{4},
J.~M.~Cordes\altaffilmark{4},
G.~H.~Janssen\altaffilmark{1},
T.~J.~W. Lazio\altaffilmark{5},
L. Petrov\altaffilmark{6},
B. W. Stappers\altaffilmark{7},
A. Lyne\altaffilmark{7}
}
\altaffiltext{1}{ASTRON, the Netherlands Institute for Radio Astronomy, Postbus 2, 7990 AA, Dwingeloo, The Netherlands}
\altaffiltext{2}{University of Wisconsin-Milwaukee, P.O. Box 413, Milwaukee, WI 53201 USA}
\altaffiltext{3}{National Radio Astronomy Observatory, Socorro, NM 87801, USA}
\altaffiltext{4}{Department of Astronomy, Cornell University, Ithaca, NY 14853, USA}
\altaffiltext{5}{Jet Propulsion Laboratory, California Institute of Technology, Pasadena, CA 91109, USA}
\altaffiltext{6}{Astrogeo Center, Falls Church, VA 22043, USA}
\altaffiltext{7}{University of Manchester, Jodrell Bank Centre for Astrophysics, Manchester M13 9PL, UK}

\begin{abstract}
Model-independent distance constraints to binary millisecond pulsars (MSPs) are of great value to both the timing observations of the radio pulsars, and multiwavelength observations of their companion stars.  Very Long Baseline Interferometry (VLBI) astrometry can be employed to provide these model-independent distances with very high precision via the detection of annual geometric parallax.  Using the Very Long Baseline Array, we have observed two binary millisecond pulsars, PSR J1022+1001 and J2145--0750, over a two-year period and measured their distances to be 700$^{+14}_{-10}$ pc and 613$^{+16}_{-14}$ pc respectively.  We  use the well-calibrated distance in conjunction with revised analysis of optical photometry to tightly constrain the nature of their massive ($M \sim 0.85$ \Msun) white dwarf companions.  Finally, we show that several measurements of their parallax and proper motion of \psrone\ and \psrtwo\ obtained by pulsar timing array projects are incorrect, differing from the more precise VLBI values by up to 5$\sigma$.  We investigate possible causes for the discrepancy, and find that imperfect modeling of the solar wind is a likely candidate for the timing model errors given the low ecliptic latitude of these two pulsars.
\end{abstract}

\keywords{astrometry --- pulsars: individual (PSR J1022+1001, J2145--0750) --- techniques: high angular resolution --- stars: white dwarfs --- stars: neutron }

\section{Introduction}

Recycled pulsars in binaries are found with companions ranging from very low mass and highly ablated objects, through ``normal" low-mass helium white dwarfs to high mass CO white dwarfs, and neutron stars.  Relatively few ``Intermediate Mass Binary Pulsars" \citep[IMBPs; e.g.,][]{camilo01a,van-kerkwijk05a} with massive CO white dwarf companions are known, and ensuring that these objects are well characterized is therefore an important aspect for understanding the different evolutionary channels that produce high-mass white dwarfs \citep{tauris12a}.

Two of the nearest and most observationally accessible IMBPs are \psrone\ and \psrtwo.
\citet{lfc96} used the \textit{Hubble Space Telescope} (\hst) to observe the
white dwarf (WD) companions of \psrone\ and \psrtwo\ and determine their effective temperatures, masses, and cooling ages.  The largest source of error in their analysis
was the pulsar distances, which they found by using the pulsar dispersion measure (DM)
and the \citet{tc93} model of the Galactic
electron density distribution.  While DM-based distances are usually
taken to have an uncertainty of around 20\%, discrepancies up to a factor of a few
are seen in some cases \citep{deller09b}.  A separate and reliable distance estimate for these two 
sources is therefore highly desirable.

Very Long Baseline Interferometry (VLBI) can provide a direct distance measurement for compact radio sources such as pulsars by measuring annual geometric parallax \citep[e.g.,][]{gwinn86a,brisken02a}.
The \psrpi\ collaboration was formed in 2010 to exploit the capabilities of the Very Long Baseline Array (VLBA) for high-precision (differential) astrometry of pulsars.

The \psrpi\ project has been enabled by two key advances: an improvement in the sensitivity of VLBI instruments, and the development of ``in-beam'' calibration techniques for differential astrometry.  VLBI astrometry can be performed in an absolute or differential fashion \citep[e.g.,][]{ma98a,reid14b}, but for radio pulsars, differential VLBI astrometry at low frequency (typically $\sim$1.5 GHz) is generally the only feasible option because almost all radio pulsars are weak and have a very steep spectrum.  The measurement accuracy for the position offset of the target from a calibrator source is affected by both the target brightness (a signal-to-noise component) and the extrapolation of the calibration from the calibrator direction to the target direction (a systematic component).  The latter term scales roughly linearly with calibrator-target angular separation \citep[e.g.,][]{chatterjee05a}.  Sensitivity improvements over the last 10 years have enabled the wide-spread use of ``in-beam" calibrators at a typical angular separation of 0.25\degrees; this greatly increases the precision of the differential measurements \citep[e.g.,][]{chatterjee09a}, but systematic errors often still dominate the error budget.  Nevertheless, with parallax accuracy at the $\sim$20$\mu$as level \citep{deller12b,deller13a}, it is feasible to measure distances with an accuracy approaching or exceeding 1\%, and in some cases it is even possible to detect the miniscule (typically $\lesssim100$ $\mu$as) orbital reflex motion of binary pulsars in the plane of the sky.

\psrpi\ has the goal of tripling the number of pulsars with precise, model-independent distance measurements and has now observed 60 pulsars with a wide variety of characteristics \citep{deller11b}.  The science motivations for the \psrpi\ project are diverse and include the refinement of models of the Galactic electron density distribution \citep{cordes02a}, the radio pulsar velocity distribution, and reference frame ties.  Based on the results being obtained from \psrpi, a subsequent project focussing specifically on millisecond pulsars, \mspsrpi, has also been initiated.  The full \psrpi\ project will be described elsewhere (Deller et al., in prep).

\begin{deluxetable*}{llccccccc}
\tabletypesize{\small}
\tablewidth{0.98\textwidth}
\tablecaption{\label{tab:targets}Observed  sources}
\tablehead{
\colhead{Source} & \colhead{Name} & \colhead{S$_{1.4}$\tablenotemark{A}} & \colhead{Separation\tablenotemark{B}} & \colhead{Spin period}       &  \colhead{Duty}  & \colhead{Orb.\ period} & \colhead{Reflex motion\tablenotemark{C}} &  \colhead{PTA\tablenotemark{D}} \\
\colhead{type}     & \colhead{}   & \colhead{(mJy)}                      & \colhead{(arcmin)}              & \colhead{ (ms)} & \colhead{cycle}                     &  \colhead{(days)}        & \colhead{($\mu$as)} & \colhead{status}
}
\startdata
Target  & PSR J1022+1001  & 6      & ---       & 16.5 & 10\% & 7.8051 & 72   &E,P   \\
Position reference       & FIRST J102310.8+100126  & 20    & 3.2     &  ---    & ---     & ----        & ---   & ---   \\
In-beam calibrator        & FIRST J102334.0+101200 & 200  & 13.5   &  ---    & ---     & ----        & ---   & ---   \\
Off-beam calibrator      & VCS1 J1025+1253 & 470  & 177.4  &  ---   & ---      & ----       & ---   & ---   \vspace{1ex}  \\ 
Target   & PSR J2145--0750 & 9      & ---        & 16.1  & 25\% & 6.8393 & 97  & E,N,P \\
Position reference       & FIRST J214557.9--074748  & 21   & 3.1       & ----    & ---      & ---  	 & ---   & ---   \\
In-beam calibrator       & FIRST J214557.9--074748  & 21   & 3.1       & ----    & ---      & ---   	 & ---   & ---   \\
Off-beam calibrator      & VCS1 J2142--0437 & 410 & 198.5   & ---     & ----     & ---  	 & ---   & ---  
\enddata
\tablenotetext{A}{Flux density (period-averaged in the case of the pulsars) at 1.4 GHz.}
\tablenotetext{B}{Angular separation from the target pulsar.}
\tablenotetext{C}{Maximum transverse displacement of the pulsar due to orbital motion (using best-fit distance and inclination obtained in Section~\ref{sec:vlbifit}).}
\tablenotetext{D}{Pulsar Timing Array status: E = EPTA pulsar, N = NANOGrav pulsar, P = PPTA pulsar.}
\end{deluxetable*}

This paper previews the results for \psrone\ and \psrtwo, which are the only binary millisecond pulsars in the \psrpi\ sample, and
Table~\ref{tab:targets} summarizes the characteristics of these two IMBPs.
To fully capitalize on the precise distance measurement provided by \psrpi\ astrometry, we also take advantage of a number of improvements in \hst\ calibration and data processing \citep{2000PASP..112.1383D, 2009PASP..121..655D} and atmosphere modeling for white dwarfs \citep{tbg11, bwd+11} to revise a number of fundamental aspects of the analysis of \citet{lfc96}.  We are able to make use of improved optical extinction values, both because of the updated distances themselves and because of significant improvements in modeling Galactic extinction \citep{2015ApJ...810...25G}.  Our new analysis, which incorporates all of the relevant uncertainties in a Markov Chain Monte Carlo (MCMC) fit, provides estimates for the temperatures, masses, and ages of the WD companions of \psrone\ and \psrtwo\ that are both more precise and more robust than previously possible.

Finally, while our primary focus for \psrone\ and \psrtwo\ is the study of their WD companions, our VLBI astrometry also has a secondary application: cross-checking the interferometric position, parallax, and proper motion against the values determined from pulsar timing to test for the presence of systematic error underestimates in either technique, and to check the alignment of the barycentric frame used for pulsar timing and the quasi-inertial reference frame used for VLBI observations.  At the present time, the relatively large uncertainty in our absolute reference positions for the two pulsars limits us to comparing only parallax and proper motion, but even these limited tests have previously been applied to just two sources \citep{deller08b,chatterjee09a}.  The timing data for both \psrone\ and \psrtwo\ is of very high quality, as both are included in Pulsar Timing Array (PTA) projects that aim to detect graviational waves in the nanoHertz regime by employing pulsars as the endpoints of a Galactic-scale gravitational-wave antenna. There are 3 active PTAs: the European Pulsar Timing Array \citep[EPTA;][]{desvignes16a} observes both pulsars, as does the Parkes Pulsar Timing Array \citep[PPTA;][]{reardon16a}, while the North American Nanohertz Observatory for Gravitational Waves \citep[NANOGrav;][]{arzoumanian15a} observes only \psrtwo.

Throughout this paper, all error bars indicate 68\% confidence intervals unless otherwise indicated.

\section{Observations and data processing}
\label{sec:obs}
\subsection{VLBI observations}
\label{sec:vlbiobs}
\begin{figure}
\begin{tabular}{c}
\includegraphics[width=0.48\textwidth]{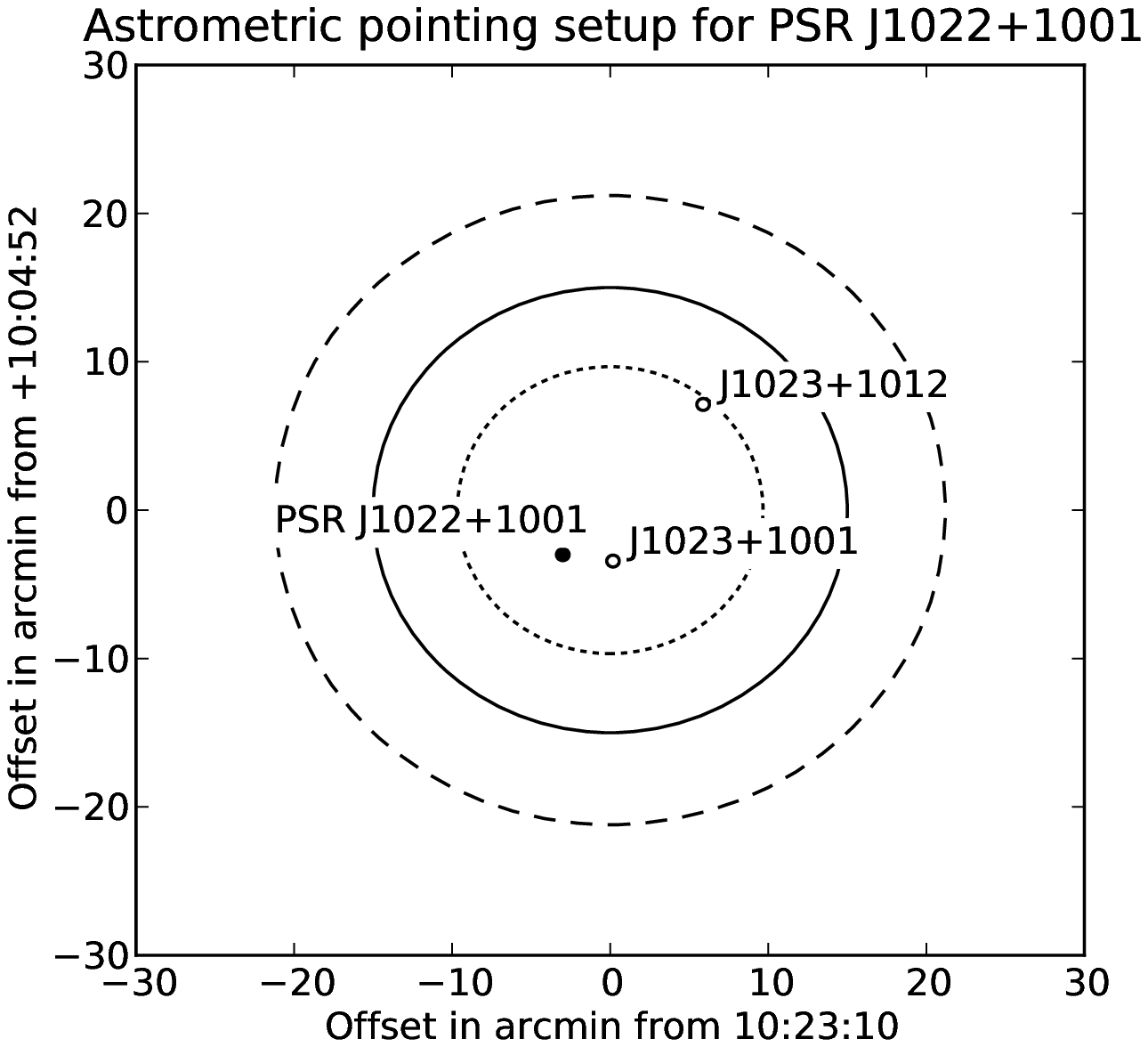} \\
\includegraphics[width=0.48\textwidth]{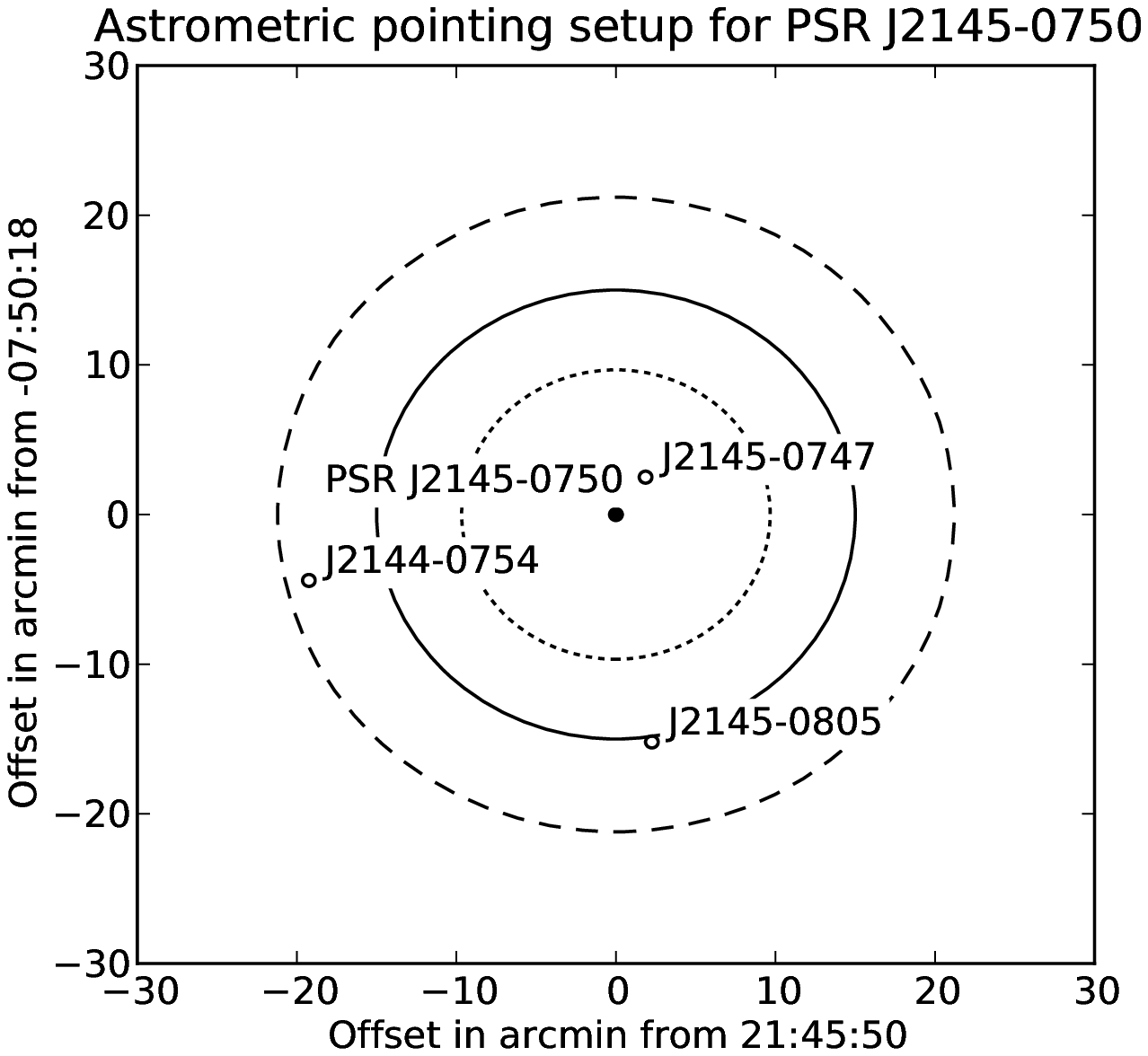}
\end{tabular}
\caption{\label{fig:pointinglayout}Location of the target pulsar (solid circle) and the selected in-beam calibrators (open circles). The dotted, solid and dashed lines show the 75\%, 50\% and 25\% response point of the primary beam at the center frequency of 1650 MHz.}
\end{figure}

Observations were made under the VLBA project code BD152.  For both \psrone\ and \psrtwo, a short observation (duration $\sim$15 minutes) was made in 2011 May to identify nearby compact radio sources that could be used as in-beam phase calibrators.  The observation setup was identical to that described for the calibrator identification program in \citet{deller13a}, with several short scans on pointing centers distributed around the target pulsar.  From each pointing, $\sim$25 correlated datasets were produced centered on known radio sources located within the VLBA primary beam, using the multi-field capability of the DiFX software correlator \citep{deller11a}.  Candidate calibrators were taken from the Faint Images of the Radio Sky \citep[FIRST;][]{becker95a} catalog. For both \psrone\ and \psrtwo, several useful calibrator sources were identified that could be observed simultaneously with the target pulsars; we refer to these from this point onwards as the ``in-beam'' calibrators.  Figure~\ref{fig:pointinglayout} shows the location of the identified in-beam calibrators, while the properties of the utilised sources are summarised in Table~\ref{tab:targets}.

Once these suitable calibrator sources were identified, nine astrometric observations were performed for each pulsar between 2011 May and 2013 June.  
In each observation, four 16 MHz dual polarization subbands spanning the frequency range 1627.49 -- 1691.49 MHz were recorded at each telescope for a total data rate of 512 Mbps per station.  All astrometric observations were made within several weeks of the parallax extrema, usually with two observations per extremum separated by 14-21 days.   A phase reference cycle time of 6.75 minutes was used, with 1 minute on the primary off-beam reference source (the sources VCS1 J1025+1253 and VCS1 J2142$-$0437) per cycle, and the remainder on the target pointing encompassing the pulsar and several in-beam calibrator sources.  For a typical epoch, with 9 VLBA antennas and 52 minutes on--target, the 1$\sigma$ image rms is $\sim$90 $\mu$Jy.  Multiple correlation passes were performed for each pulsar for each epoch, including one pass at the position of each in-beam calibrator, one pass at the position of the target pulsar without any special processing (the ``ungated'' pass), and one pass at the position of the target pulsar using the pulsar gating capability of the DiFX correlator \citep{deller07a} to improve the S/N (the ``gated" pass).  The pulsar ephemeris and gate parameters for the gated pass were generated using timing information from the Lovell telescope at the Jodrell Bank Observatory in the UK, which observes \psrone\ and \psrtwo\ every few weeks.

\subsection{VLBI data processing}
\label{sec:vlbiproc}
The data reduction used was very similar to that presented in \citet{deller13a} and will be described in detail in a forthcoming \psrpi\ catalog paper (Deller et al. in prep); we summarize it briefly here.  We made use of the ParselTongue \citep{kettenis06a} Python interface to the AIPS package \citep{greisen03a}.  For each observation, standard on-source flags were applied, followed by standard amplitude calibration (AIPS tasks ACCOR and APCAL) and corrections accounting for updated Earth orientation parameters and dispersive delays estimated from global ionospheric models  (AIPS tasks CLCOR and TECOR).  Further amplitude corrections for the antenna primary beam response were then applied, using the same custom ParselTongue script described in \citet{deller14a}.  Bandpass calibration using an amplitude calibrator source and delay/phase calibration using the primary off-beam phase reference source (AIPS tasks BPASS and FRING) followed.  The amplitude scale was refined using amplitude self-calibration on the phase reference source (AIPS task CALIB) with a solution interval of 20 minutes.  All steps made use of calibrator structure models derived by imaging a concatenated dataset formed from all 9 astrometric epochs.

The amplitude, delay, and phase calibration so derived was then transferred to the target field, allowing us to obtain images of the primary in-beam calibrators relative to the off-beam calibrators.  Details of the in-beam calibrators are given in Table~\ref{tab:targets}.  As with the off-beam sources, a model of each in-beam calibrator was formed by imaging a concatenated dataset from all 9 epochs.  We used these models to perform phase self-calibration on the primary in-beam calibrator with a solution interval of 10 seconds (for FIRST J102334.0+101200) or 20 seconds (for FIRST J214557.9--074748).  These phase refinements were then applied to the target pulsars and other in-beam calibrator sources, meaning each of these sources could be imaged to give a position relative to the primary in-beam calibrator.  Finally, \psrtwo\ showed substantial intra-observation amplitude variations attributable to diffractive scintillation (the predicted diffractive scintillation bandwidth is $\gtrsim$100 MHz at our observing frequency), and so we derived and applied per-subband amplitude corrections on a timescale of 10 minutes for \psrtwo\ only, using the procedure described in \citet{deller08a}.  The calibrated data was then split, divided by the model of calibrator structure in the case of in-beam calibrator sources, and exported from AIPS for imaging in the Difmap package \citep{shepherd97a}.

We extracted the astrometric observables (position offsets from the primary in-beam calibrators) in the image plane for the target pulsars and the in-beam calibrator sources.  Imaging was performed in an automated fashion on the entire available bandwidth, using a model initialised with a point source placed at the peak of the dirty image, followed by a 50-iteration model fit.  The resultant Stokes I clean image was written to disk in FITS format.  We also divided the datasets in half in time and repeated this process on the first and second half, generating two additional images per observation that we used to estimate the systematic astrometric errors as described below.  The clean images were loaded into AIPS and the source position (and errors on the position parameters) were estimated using the gaussian fitting task JMFIT.  

\subsection{Analysis of the position time series}
\label{sec:vlbifit}
We repeated the procedure in Section~\ref{sec:vlbiproc} for each of the nine astrometric epochs.  The resultant dataset comprised a time series of nine astrometric positions for the target pulsars and for the in-beam calibrators, referenced to the assumed position of the primary in-beam calibrator source.  For \psrone, the fainter calibrator source FIRST J102310.8+100126 was considerably closer to the pulsar than the bright primary in-beam calibrator FIRST J102334.0+101200, so we subtracted the position residuals for FIRST J102310.8+100126 from those for \psrone\ to make it the effective position reference and reduce the systematic errors in the pulsar position time series caused by residual ionospheric errors.  The position residuals for FIRST J102310.8+100126 were consistent with expectations for differential astrometry over a separation of $\sim$12\arcmin, with an rms scatter of 80 $\mu$as in right ascension and 190 $\mu$as in declination. For \psrone, subtracting the FIRST J102310.8+100126 residuals approximately halved the rms scatter in the post-fit residuals in each coordinate to $\sim$100\,$\mu$as, which highlights the benefit of the fourfold reduction in the angular separation between target and position reference. For \psrtwo, FIRST J214557.9--074748 was already the closest in-beam calibrator and so this step was unnecessary.

For each pulsar, we could now fit astrometric parameters to the obtained position offsets with respect to the nearest in-beam calibrator.  Since these are both binary pulsars, in addition to the five usual astrometric parameters (position in right ascension and declination, proper motion in right ascension and declination, and parallax) our fit also included the binary inclination $i$ and longitude of ascending node $\Omega$ to account for the pulsar reflex motion.  We limited the allowable range of $i$ assuming a companion mass $0.5 \Msun < \Mc < 1.35 \Msun$ and pulsar mass of $1.2 \Msun < \Mp < 2.4 \Msun$, which gives $i>26\degrees$ for \psrone\ and $17\degrees < i < 55\degrees$ for \psrtwo.  As we see in Section~\ref{sec:discussion}, these restrictions are considerably looser than the limits we can ultimately derive on $i$ based on the optical modeling, and are consistent with the marginal detection of Shapiro delay for \psrone\ \citep{reardon16a}. The remaining Keplerian binary parameters were fixed using the pulsar timing ephemerides provided in \citet{reardon16a}.

Both pulsars also have significant measurements of $\dot{x}$, the apparent time rate of change of the projected semi-major axis, which is dominated by a kinematic term dependent on $i$, $\Omega$, and the pulsar proper motion \citep{kopeikin96a}. Accordingly, large areas of $\left( \Omega, i \right)$ space are excluded by the combination of $\dot{x}$ from pulsar timing and the VLBI proper motion.  Independent datasets from all three PTAs give consistent $\dot{x}$ for \psrtwo: \citet{reardon16a} gives $(8.0 \pm 0.8) \times 10^{-15}$ for the PPTA, \citet{desvignes16a} gives $(8.2 \pm 0.7) \times 10^{-15}$ for the EPTA, and \citet{fonseca16a} gives $(10 \pm 2) \times 10^{-15}$ for NANOGrav.  \psrone, on the other hand, is observed only by PPTA and EPTA, which give inconsistent results: $(1.15 \pm 0.16) \times 10^{-14}$ and $(1.79 \pm 0.12) \times 10^{-14}$  respectively.  Despite the large discrepancy, we found the impact on the astrometric fit to be relatively small; the results presented below use the PPTA $\dot{x}$ constraints from \citet{reardon16a} for both pulsars as the primary constraint on $i$ and $\Omega$ in the astrometric fitting process, and we examine the effect of the discrepancy for \psrone\ in Section~\ref{sec:vlbivstiming}.

The formal position errors we have available are underestimates of the true error, since they do not include systematic position shifts due to the residual ionospheric calibration errors between the nearest in-beam calibrator and the target pulsar.  A least-squares fit is therefore a poor way to estimate the errors on the fitted parameters -- the reduced $\chi^2$ of the fit is considerably larger than unity, and the errors on the fitted parameters would be under-estimated.  Instead, we follow the recent practice \citep[e.g.,][]{chatterjee09a,deller13a} of using a bootstrap fit \citep{efron91a}: we ran 50,000 trials, where in each trial we selected {\em with replacement} 9 position measurements from the pool of 9 epochs, and performed a least-squares fit to the selected positions.  Taking the resultant 50,000 values for each of the fitted parameters (reference position, proper motion, parallax, $i$, and $\Omega$), we form a cumulative probability density function and extract the most probable value and 68\% confidence interval, which we report in Table~\ref{tab:fit}. The bootstrap--derived errors are more conservative (by a factor of a few) than a straightforward least squares fit to the 9 epochs using formal position errors only.

To verify the robustness of these error estimates, we investigated several alternatives in which we attempt to estimate the systematic error contributions.  First, we compared the position differences between the two halves of a single observation, and added an error term in quadrature to each epoch which was proportional to this apparent intra-epoch shift.  Second, we added a constant error term in quadrature to all epochs based on the scatter in the position residuals from a least-squares fit to the 9 epochs.   The per-epoch systematic error estimated in this way ranged from 50 --150 $\mu$as per epoch in both right ascension and declination.  If this estimate of systematic error is added to the formal position errors and the bootstrap repeated, then results which are consistent to well within 1$\sigma$ are obtained, but with slightly smaller errors on the fitted parameters.  This somewhat counter-intuitive result can be understood by considering the effect of the handful of position measurements with very small formal errors on the bootstrap -- the presence or absence of these points in a given trial can considerably change the fitted parameters for that trial.  If all position measurements have an error floor due to the estimated systematic error contribution, the leverage of the points with small formal errors is considerably reduced, reducing the variations between trials.  Even though there is good justification for adding an estimated systematic error contribution to the position errors, we report the results of the bootstrap in which no estimate of systematic error added to the position fits, as this gives the most conservative results.  

In addition to fitting the pulsar motion with respect to the nearest in-beam calibrator, we also measured the absolute position uncertainty of the pulsar reference position.  The absolute position uncertainty contains contributions from the off-beam calibrator reference position ($\ll1$ mas), the frequency-dependent shift in the off-beam calibrator position ($\sim$1 mas), and the offset of the primary in-beam calibrator from the off-beam calibrator.  This last term typically dominates, and was estimated to be $\sim$2 mas by measuring the scatter of the position of the in-beam calibrator in images where no self-calibration was applied.

\begin{deluxetable*}{lrrr}
\tabletypesize{\small}
\tablewidth{0.9\textwidth}
\tablecaption{\label{tab:fit}Fitted astrometric parameters for \psrone\ and \psrtwo.}
\tablehead{
\colhead{Parameter} & \colhead{\psrone} & \colhead{\psrtwo} 
}
\startdata
Right ascension (J2000)\tablenotemark{a}
							& 10:22:57.9957(1)
							& 21:45:50.4588(1) \\
Declination (J2000)\tablenotemark{a}
							& 10:01:52.765(1)
							& $-$07:50:18.513(2) \\
Right ascension offset (mas)\tablenotemark{b}
							& $-$190721.06(3)
							& $-$111530.71(3) \\
Declination offset (mas)\tablenotemark{b}
							& 25883.74(3)
							& $-$149818.42(7) \\
Position epoch (MJD)			& 56000 & 56000 \\
$\mu_{\alpha}$	(mas yr$^{-1}$)		& $-$14.86 $\pm$ 0.04 
							& $-$9.46 $\pm$ 0.05 \\
$\mu_{\delta}$	(mas yr$^{-1}$)		& 5.59 $\pm$ 0.03	
							&  $-$9.08 $\pm$ 0.06 \\
Parallax (mas)	 				& 1.43$^{+0.02}_{-0.03}$
							& 1.63 $\pm$ 0.04  \\
Distance (pc)					& 700$^{+14}_{-10}$
							& 613$^{+16}_{-14}$ \\
$v_{\mathrm T}$ (km s$^{-1}$)		& 52.6$^{+1.3}_{-0.9}$
							& 38.1$^{+1.2}_{-1.1}$ \\
$\Omega$ (\degrees)\tablenotemark{c}
							& 336$^{+12}_{-36}$
							& 220 $\pm$ 12 \\
$i$ (\degrees)\tablenotemark{d}	
							& 42$^{+20}_{-16}$
							& 21$^{+7}_{-4}$ \\
Reflex motion amplitude ($\mu$as)	& 72$^{+38}_{-19}$  
							& 97$^{+22}_{-23}$
\enddata
\tablenotetext{a}{The errors quoted here are dominated by the estimated absolute position uncertainty transferred from the in-beam calibrators.}
\tablenotetext{b}{The offset of the pulsar from the reference in-beam calibrator position.}
\tablenotetext{c}{Measured clockwise (towards North) from East \citep{kopeikin96a}.}
\tablenotetext{d}{When considering orbital motion in the radial direction only (as is typical for pulsar timing), then the inclination range $180\degrees > i > 90\degrees$ is degenerate with $0\degrees < i < 90\degrees$, and $i$ is usually only quoted in the range 0 -- 90\degrees.  Fitting for reflex motion allows us to distinguish between $0\degrees < i < 90\degrees$ and $90\degrees < i < 180\degrees$, but here we quote $i$ folded into the range 0 -- 90\degrees.}
\end{deluxetable*}

\begin{figure}
\begin{center}
\begin{tabular}{c}
\includegraphics[width=0.49\textwidth]{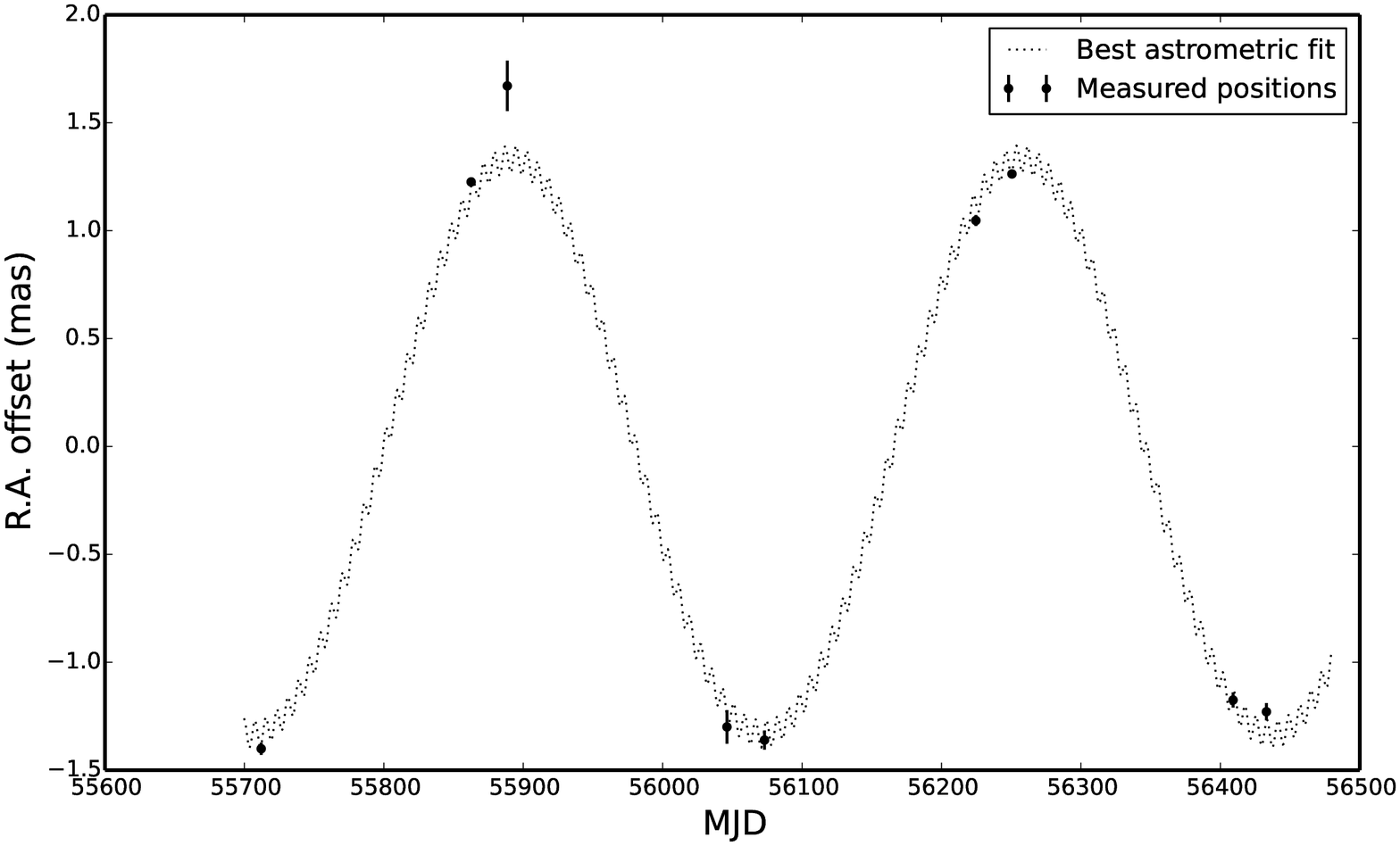} \\
\includegraphics[width=0.49\textwidth]{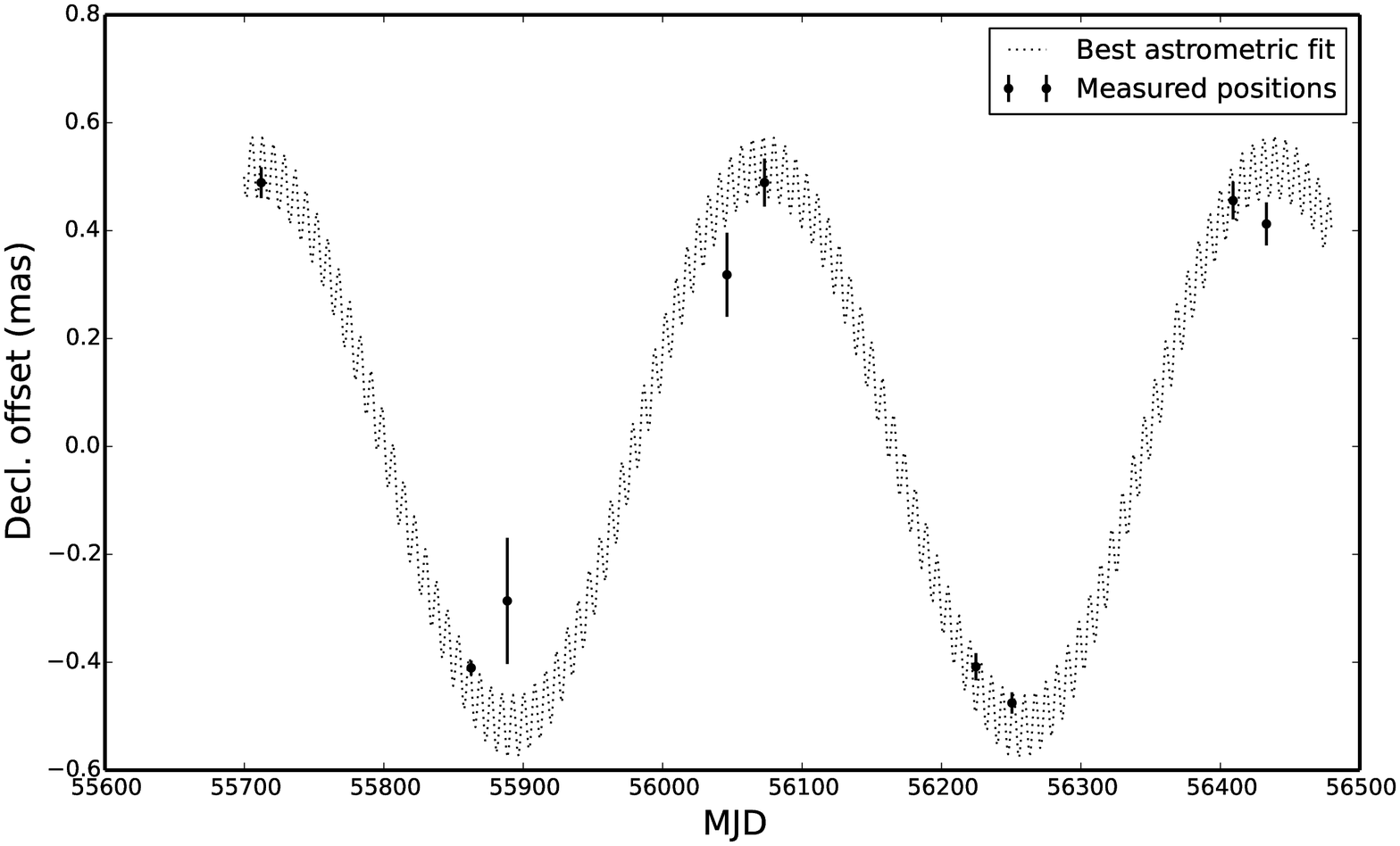} 
\end{tabular}
\end{center}
\caption{\label{fig:fit1}The best astrometric fit for \psrone, with best-fit proper motion subtracted to highlight the parallax and reflex motion.  The top panel and bottom panels shows offset from the reference position (at MJD 56000) in right ascension and declination respectively. Error bars show formal position fit errors that underestimate the position error at each epoch, particularly when the pulsar is bright and strongly detected, but the bootstrap approach means that the errors on the output parameters of interest (parallax and proper motion) are not underestimated as a result.}
\end{figure}

\begin{figure}
\begin{center}
\begin{tabular}{c}
\includegraphics[width=0.49\textwidth]{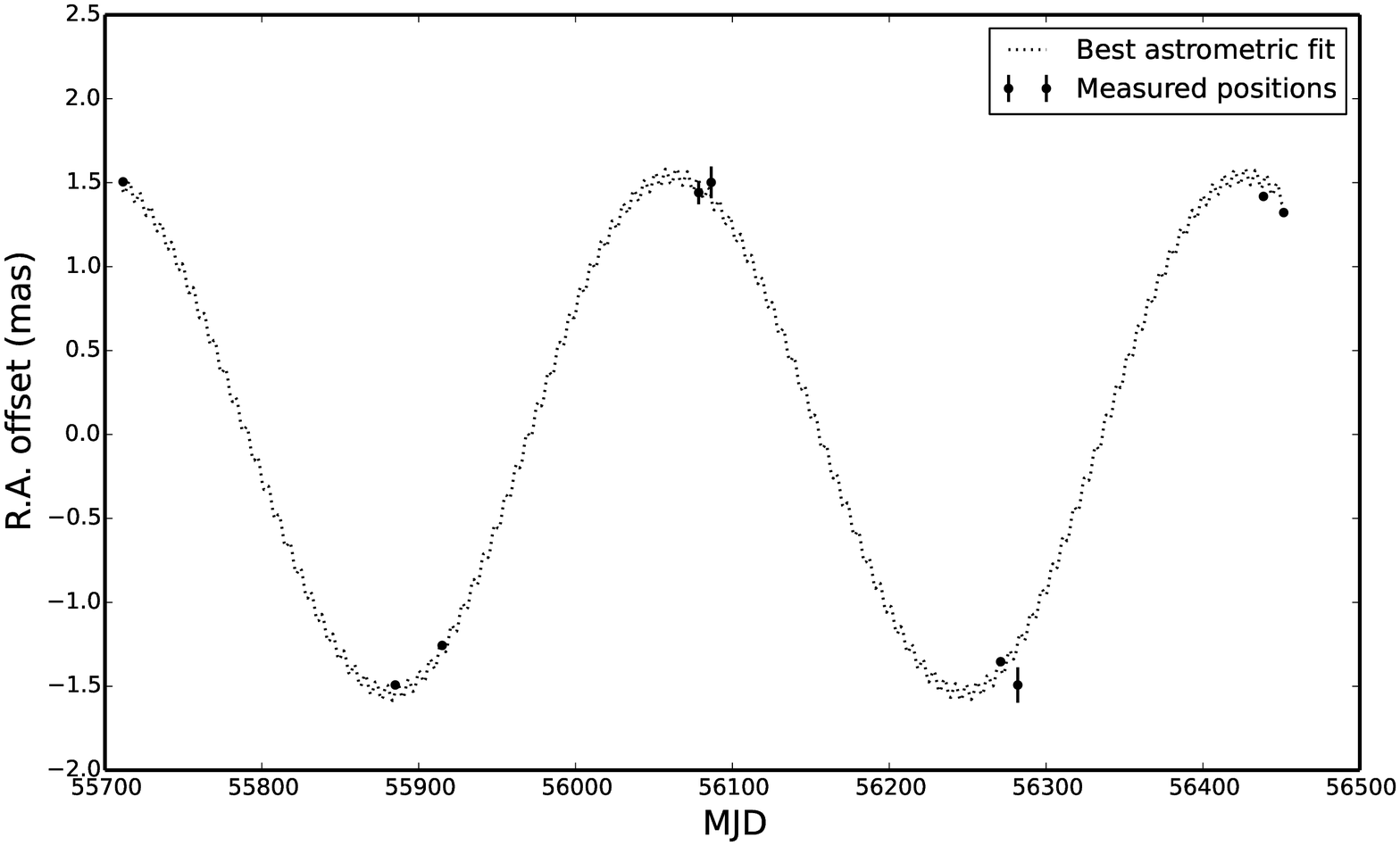} \\
\includegraphics[width=0.49\textwidth]{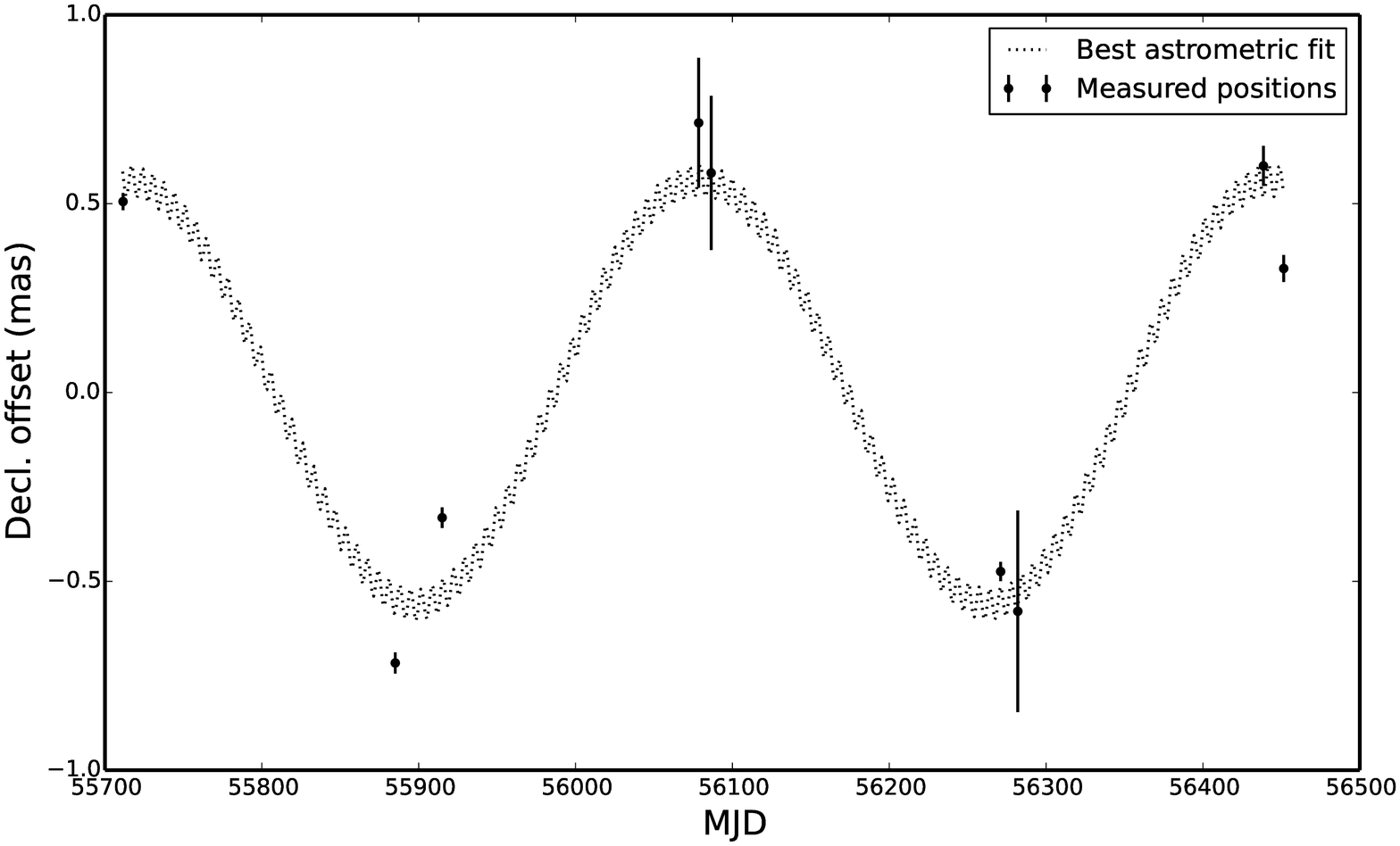} 
\end{tabular}
\end{center}
\caption{\label{fig:fit2}The best astrometric fit for \psrtwo, with best-fit proper motion subtracted to highlight the parallax and reflex motion.  The top panel and bottom panels shows offset from the reference position (at MJD 56000) in right ascension and declination respectively. Error bars show formal position fit errors that underestimate the position error at each epoch, particularly when the pulsar is bright and strongly detected, but the bootstrap approach means that the errors on the output parameters of interest (parallax and proper motion) are not underestimated as a result.}
\end{figure}

Figures~\ref{fig:fit1} and \ref{fig:fit2} show the measured positions for \psrone\ and \psrtwo\ respectively, along with the best fit model.

\subsection{Optical observations}
\label{sec:opticalobs}
Observations were made with the Wide Field and Planetary Camera 2 (WFPC2) 
aboard \hst\ between 
1995-June and 1995-October.  Both sources were observed in the 
F555W ($V$-band) and F814W ($I$-band) filters, while \psrtwo\ was 
also observed in the F439W ($B$-band) filter. The observations 
consisted of two identical exposures for each target in each band 
with no dithering between them.  The individual exposure times 
varied from 1100\,s for the F439W filter to 400\,s for the F814W
filter.  The pulsars were all on the Planetary Camera (PC) detector,
and we only analyzed those images.
\begin{deluxetable}{c c c}
  \tablewidth{0pt}
\tablecaption{Optical Photometry\label{tab:photometry}}
\tablehead{
  \colhead{Parameter} & \colhead{\psrone} & \colhead{\psrtwo}}
\startdata
$A_V$\tablenotemark{a} & $0.07\pm0.09$ & $0.09\pm0.09$ \\
$m_{\rm F439W}$ (mag) & \nodata & $24.20\pm0.07$ \\
$m_{\rm F555W}$ (mag) & $23.07\pm0.02$ & $23.67\pm0.03$ \\
$m_{\rm F814W}$ (mag) & $22.64\pm0.04$ & $22.97\pm0.05$
\enddata
\tablecomments{All photometry is in the Vega system.}
\end{deluxetable}

\begin{figure}
\begin{center}
  \includegraphics[width=0.9\columnwidth]{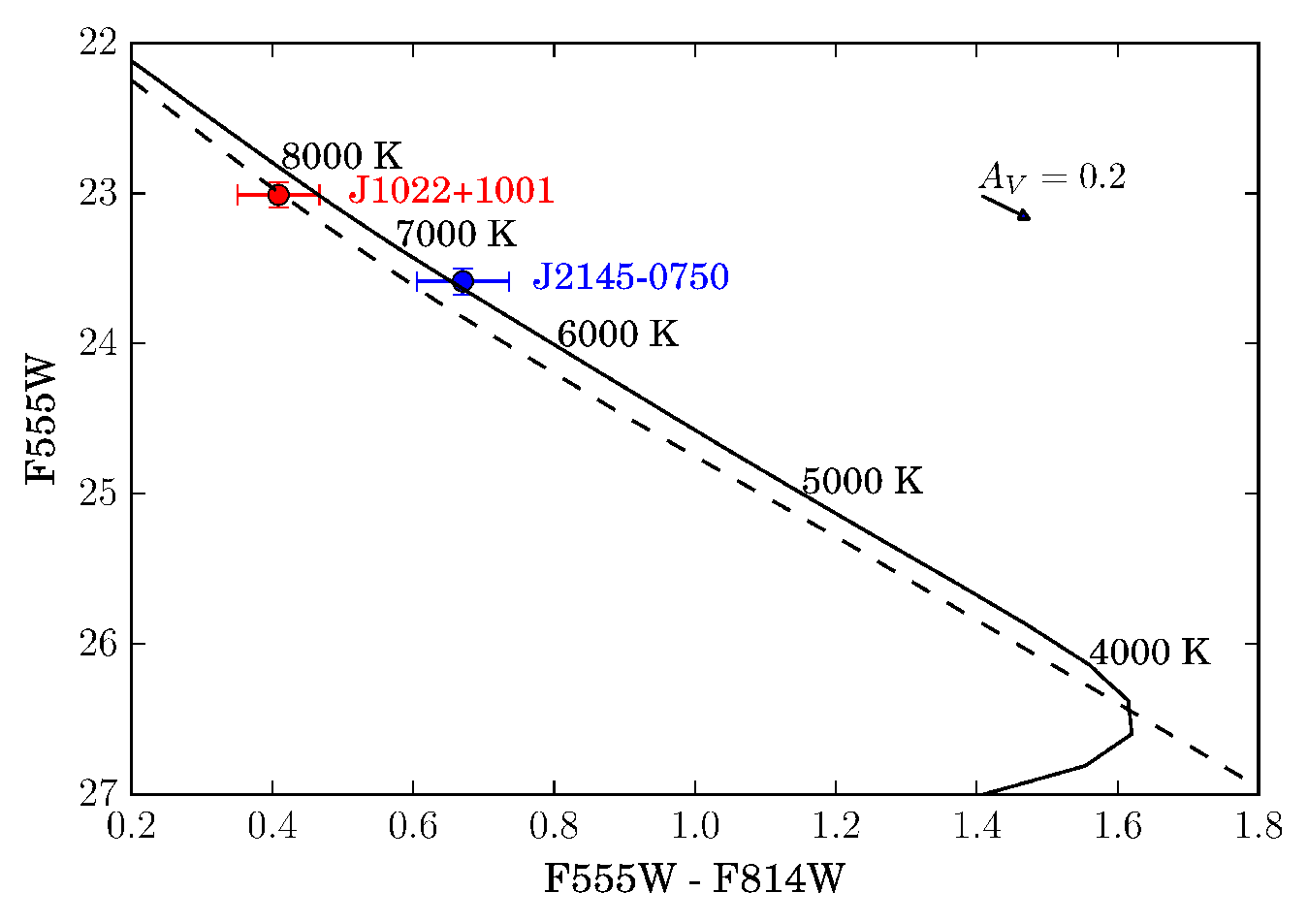}
  \caption{\label{fig:cmd}Color-magnitude diagram. 
    The red and blue filled circles with error bars indicate the \hst\ data 
    for \psrone\ and \psrtwo\, respectively.  
    We also show unreddened model tracks for $0.9 M_\odot$ WDs with 
  hydrogen (solid) and helium (dashed) atmospheres at a distance of
  650\,pc, with effective temperatures labeled. The arrow shows a reddening 
  vector for $A_V=0.2\,$mag.}
  \end{center}
\end{figure}

Table~\ref{tab:photometry} gives the photometry data for \psrone\ and 
\psrtwo.  To compute the extinction $A_V$, we used the VLBI distances 
from Table~\ref{tab:fit} to infer the 
reddening $E(B-V)$ along the line-of-sight based on the three-dimensional
dust model from \citet{2015ApJ...810...25G}.    The reddening was
converted into extinctions in each band using $R_V=3.1$ and the
extinction coefficients $A_\lambda$ from \citet[][for
  $\Teff=5000\,$K]{2008PASP..120..583G}, although we reduced each
$A_\lambda$ by 15\% to account for the revised calibration of \citet{2011ApJ...737..103S}. 
Data reduction followed the recommendations for \texttt{HSTphot}
\citep[version 1.1;][]{2000PASP..112.1383D}: we masked bad pixels,
estimated the sky level, and masked cosmic rays through comparison of
exposure pairs.  We performed point-spread function (PSF) fitting
photometry using the revised calibration of
\citet{2009PASP..121..655D}.  The fields were rather sparse so we did
not allow \texttt{HSTphot} to refine the PSFs or aperture corrections
but just used the nominal values. We tested the robustness of our 
data reduction technique by varying the preprocessing steps and PSF 
fitting options, and found no significant variations in the magnitudes.  
Figure~\ref{fig:cmd} plots the photometry on a color-magnitude 
diagram, where the filled circles with error bars represent the \hst\ data, and 
curves show models for
H-atmosphere (DA) and He-atmosphere (DB) WDs, 
labeled with effective temperature.

\begin{figure}
	\begin{tabular}{c}
	\includegraphics[width=0.49\textwidth]{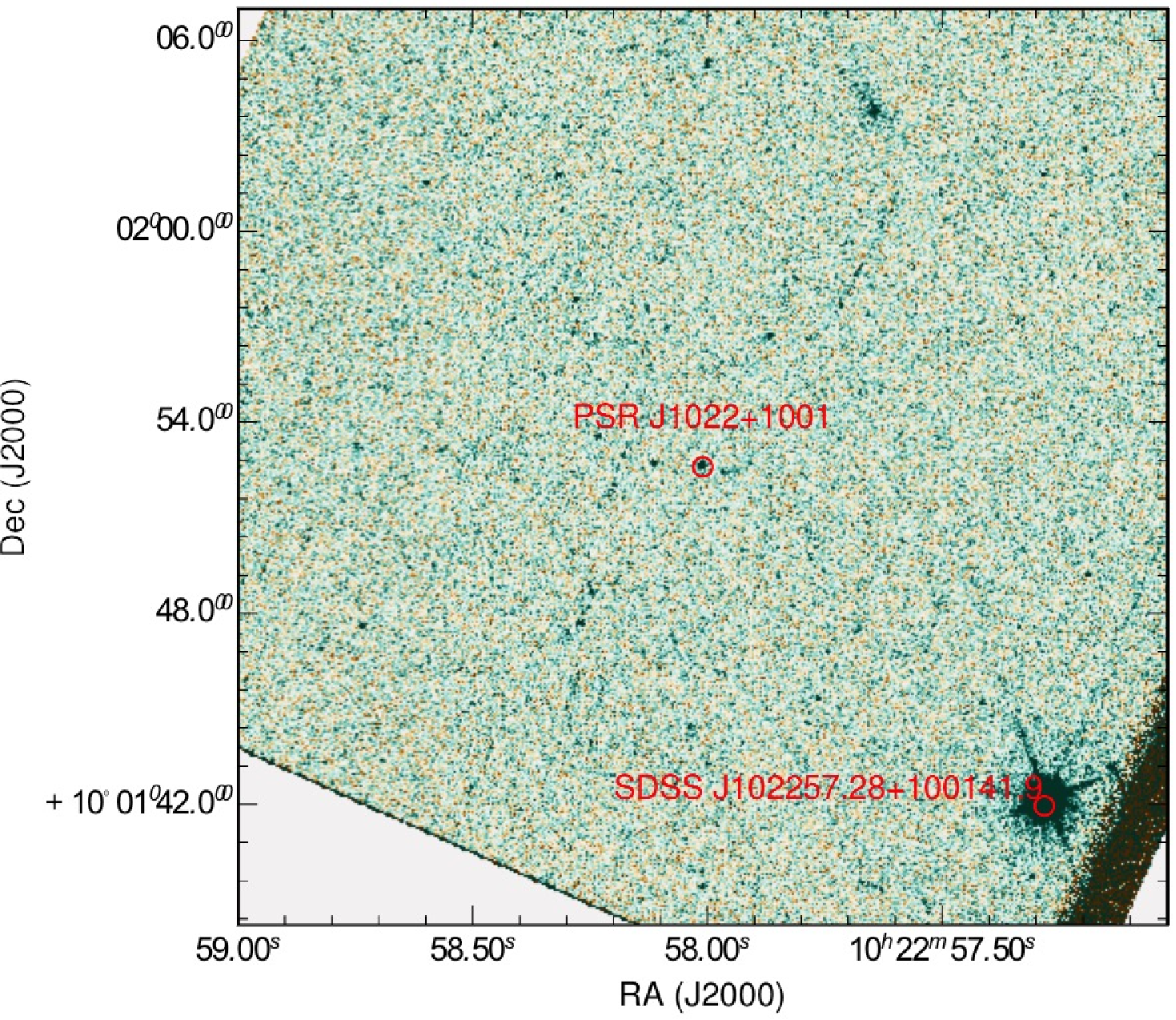} \\
	\includegraphics[width=0.49\textwidth]{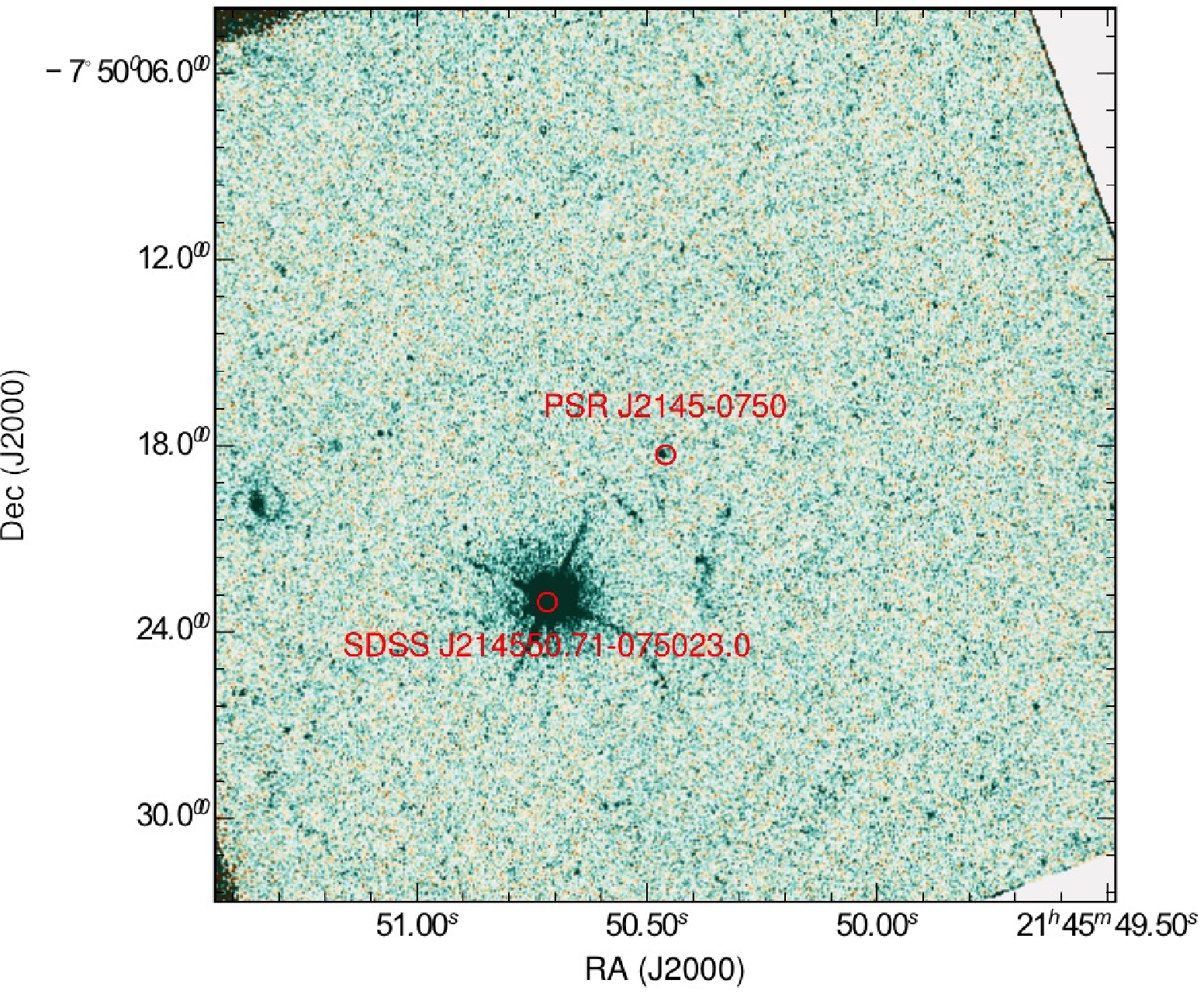}
	\end{tabular}
	\caption{\label{fig:HSTimages}Drizzled \citep{2002PASP..114..144F} \hst/WFPC2 images of \psrone\ and \psrtwo.  In each case, the central circle is centered on the VLBI position of the pulsar corrected to the epoch of the observations and has a radius of $0\farcs3$, consistent with the absolute astrometric uncertainty of the Hubble Legacy Archive reprocessing.  We also show an additional reference source from the Sloan Digital Sky Survey \citep{aaa+14} that confirms the absolute astrometric accuracy.  In both cases the pulsar's position is consistent with the proposed optical counterpart \citep[cf.][]{lfc96}.}
\end{figure}

We also downloaded drizzled combined images for each filter from the
Hubble Legacy Archive.  These images have had their absolute
astrometry improved to $0\farcs3$ on average, which we confirm through
examination of reference sources.  As seen in Figure~\ref{fig:HSTimages},
in both cases the optical counterpart is within $0\farcs3$ of the
pulsar position, which is a significant improvement on the astrometry 
in \citet{lfc96}.

\section{Discussion}
\label{sec:discussion}
\subsection{The distance to \psrone\ and \psrtwo\ and kinematic corrections to observed timing parameters}
The distances provided by our precise measurement of parallax for \psrone\ and \psrtwo\ allows us to evaluate the correctness of previous distance estimates for these pulsars.  Since the significance of the parallax measurement is so high, bias corrections when converting the measured VLBI parallax to an estimated distance are negligible \citep[e.g.,][]{verbiest12a}. 

A distance estimate based on the pulsar DM and the NE2001 model of the Galactic electron density distribution \citep{cordes02a} is typically assumed to have an error of $\sim$20\%, although individual pulsars can often differ by far more than this amount \citep[e.g.,][]{deller09b}.  This is the case for \psrone, where the parallax distance of 0.7 kpc is over 50\% larger than the NE2001 distance of 0.45 kpc.  The NE2001 distance for \psrtwo\ (0.57 kpc), on the other hand, is consistent at the 10\% level with our parallax distance.  For both pulsars, timing parallax results with a 10--20\% precision were available; as we show in Section~\ref{sec:vlbivstiming}, the EPTA measurement for \psrone\ is incorrect at the 3.5$\sigma$ level.  Thus, for \psrone\ in particular, using either DM or timing based distance estimates would have substantially biased the WD modeling presented in Section~\ref{sec:wdmodel} below.

We can also use the VLBI distance and transverse velocity to calculate kinematic contributions to the observed pulsar spin period derivative and orbital period derivative \citep{shklovskii70a,damour91a}.  For \psrone\ we find $\dot{P}_{\mathrm{kin}} = (4.9\pm0.2)\times10^{-21}$, or 11\% of the observed $\dot{P}$, while for \psrtwo\ we find $\dot{P}_{\mathrm{kin}} = (2.9\pm0.2)\times10^{-21}$, or 10\% of the observed $\dot{P}$.  Of the two pulsars, only \psrone\ has a measured value of $\dot{P_b}$ \citep[$(5.5\pm2.3)\times10^{-13}$;][]{reardon16a}, which is consistent at the $\sim$1.5$\sigma$ level with our calculated value of $\dot{P_b}_{\mathrm{kin}} = (2.0\pm0.1)\times10^{-13}$.

\subsection{Modeling the WD companions of \psrone\ and \psrtwo}
\label{sec:wdmodel}
We fit the photometry given in Table~\ref{tab:photometry} using 
both hydrogen (DA) and helium (DB) models from \citet{tbg11} and \citet{bwd+11}, 
respectively\footnote{Also see \url{http://www.astro.umontreal.ca/{\til}bergeron/CoolingModels/}.}.
These models tabulate synthetic photometry integrated throughout the
\hst/WFPC2 filter passbands in the Vega system (like
\texttt{hstphot}) for a range of effective temperatures and masses
(and hence ages or surface gravities). We only use those for masses
$\geq 0.4\,M_\odot$ where the assumption of a carbon/oxygen core (as
opposed to a helium core) is likely correct. We use the affine invariant Markov chain
Monte Carlo (MCMC) ensemble sampler \texttt{emcee} \citep{fmhlg13} to perform the 
model fits.

The companions to \psrone\ and \psrtwo\ are both warm, massive WDs, with masses $\sim 0.8\,M_\odot$ 
and effective temperatures $\sim 6000 - 9000\,\mathrm{K}$.  
Since the variation of WD color with mass is small over that temperature 
range, we simplified our analysis by using the colors for a single WD model 
(specifically 0.8\,$M_\odot$), adjusted the absolute magnitudes 
$M$ for a radius of $0.01\,R_\odot$ (which is the radius of a 0.8\,$M_\odot$ WD),
and computed the apparent model magnitude $m$ according to:
\[
m = M +5 \log_{10}\left(\frac{d}{10\,{\rm pc}}\right) + A_\lambda A_V -
  5\log_{10} \left[\frac{\Rc(\Teff,\Mc)}{0.01\,R_\odot}\right]
  \]
This simplification introduces an error which is less than $0.01\,$mag over most of the parameter space, 
less than the uncertainty of the absolute 
calibration of both \texttt{hstphot} and the synthetic photometry, 
which is roughly 1\%.  The error can exceed 0.01\,mag (as high as 0.03\,mag) only for
very high or very low mass WDs, which are strongly disfavoured, and even then it is much smaller
than the measurement uncertainty.

The independent variables are the effective temperature \Teff, 
the distance $d$, the $V$-band extinction $A_V$, 
the WD mass \Mc, and the binary inclination $i$. 
Note that it is possible to describe a WD without
reference to inclination, but including this parameter enables us
to reject unphysical combinations of the WD and pulsar masses.  
The inclination, pulsar mass, and WD mass are related via the 
mass function:
\[
\frac{(\Mc \sin i)^3}{(\Mc+\Mp)^2} = \frac{4\pi^2 x^3}{T_\odot P_b^2}
\]
with $x$ the projected semi-major axis of the pulsar's orbit, $P_b$
the binary period, and $T_\odot=GM_\odot/c^3$.  In our analysis, we use 
the mass function to eliminate all combinations of $\Mc$ and $i$ that result 
in a pulsar mass below $1.2\,M_\odot$; this lower limit is based on the 
observed distribution of pulsar masses \citep{2012ApJ...757...55O}.  
Additionally, we use the distributions for $\Mc$ and $i$ to compute the 
distribution of pulsar mass $\Mp$, using orbital parameters obtained via 
pulsar timing \citep{reardon16a}.

Our priors on the parameters are:
\begin{description}
\item[\Teff] Uniform over the model grid.
\item[$d$] Gaussian taken from the values in Table~\ref{tab:fit}.
\item[$A_V$] Gaussian taken from the values in
  Table~\ref{tab:photometry}, required to be $\geq0$.
\item[\Mc] Uniform over the model grid, except that we also require
  \Mc\ to be greater than the minimum companion mass determined 
  from $P_b$ and $x$, assuming a (conservative) 
  neutron star mass of $\Mp=1.2\,M_\odot$.
\item[$i$] Uniform in $\cos i$, except that we reject
  combinations of $\Mc$ and $i$ that result in pulsar masses $<1.2\,M_\odot$ 
  or $> 2.4 \,M_\odot$, and reject values of $i$ excluded at $>$3$\sigma$ by $\dot{x}$ constraints
  provided by pulsar timing.  For \psrone, where two recent measurements
  of $\dot{x}$ are inconsistent \citep{reardon16a,desvignes16a} we use the 
  PPTA measurement \citep{reardon16a}, which is the most permissive.
\end{description}

The radius and age were computed as a function of both mass and effective
temperature using a bilinear interpolation over the model values.  The
apparent magnitude in each band was compared against our observations
to determine the posterior pdf. We started the MCMC with 100 ``walkers'' and 
iterated it for 5000 iterations. We ignored the first 50 samples to
account for burn-in, and we thinned the results by a factor
of 51 to account for correlations between adjacent samples \citep{fmhlg13}.

\begin{figure*}
	\begin{center}
	\includegraphics[width=0.9\textwidth]{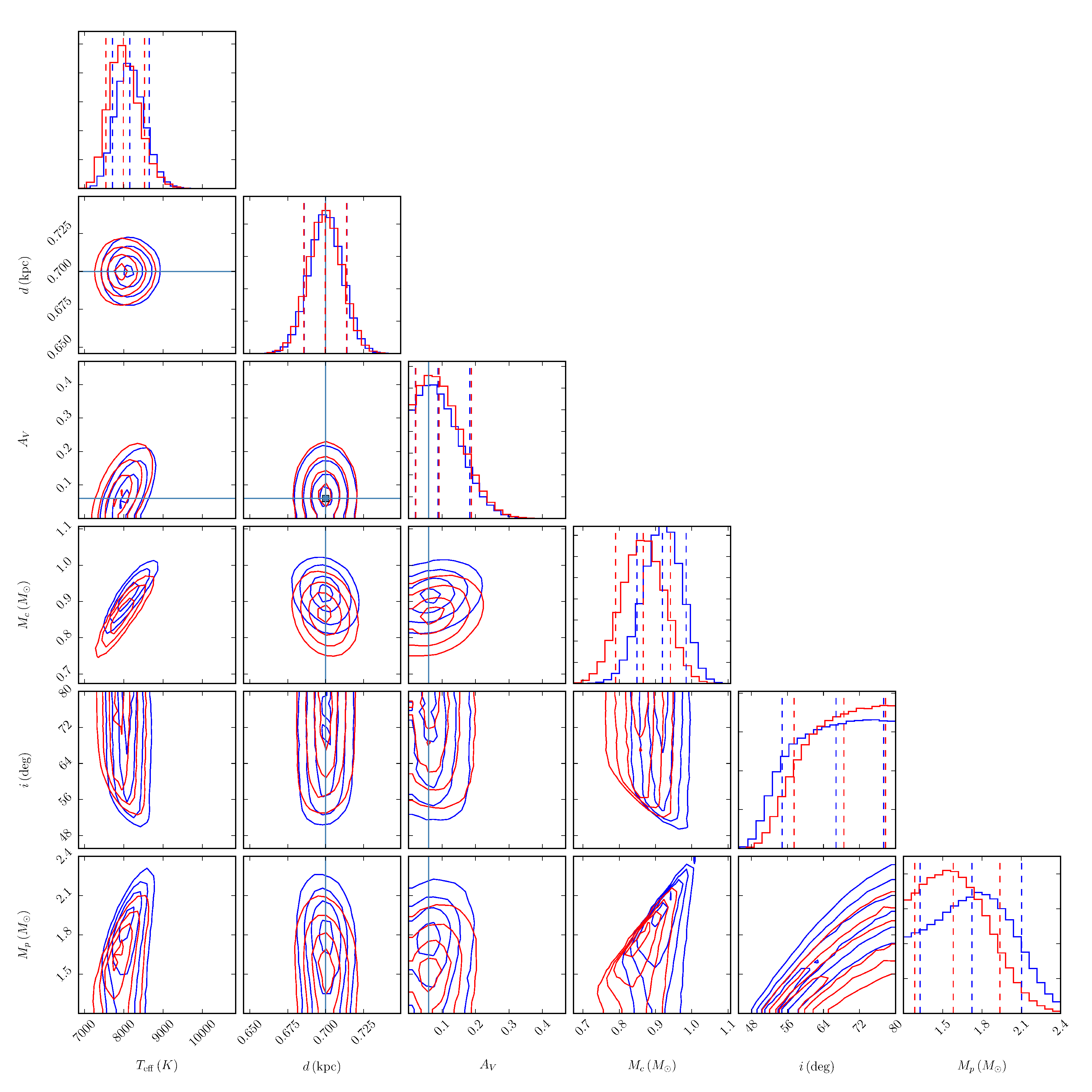}
	\caption{\label{fig:phot_fit1022}Joint two-dimensional posterior probability distribution functions and 
		marginalized one-dimensional posterior probability distribution functions 
		from the MCMC analysis on \psrone.  The prior distributions are as 
		described in Sec.~\ref{sec:discussion}.  These results used the synthetic 
		photometry for a WD mass of 0.8\,$M_\odot$, but changing that to 0.7 or 
		0.9\,$M_\odot$ resulted in identical constraints.  The blue curves are for 
		hydrogen-atmosphere (DA) WDs, while the red curves are for 
		helium-atmosphere (DB) WDs.  The vertical dashed lines show the 
		median, 10\%, and 90\% confidence limits.  The horizontal/vertical solid 
		lines show the nominal values of $d$ and $A_V$ from Table~\ref{tab:fit} 
		and \ref{tab:photometry}, respectively.}
	\end{center}
\end{figure*}

\begin{figure*}
	\begin{center}
	\includegraphics[width=0.9\textwidth]{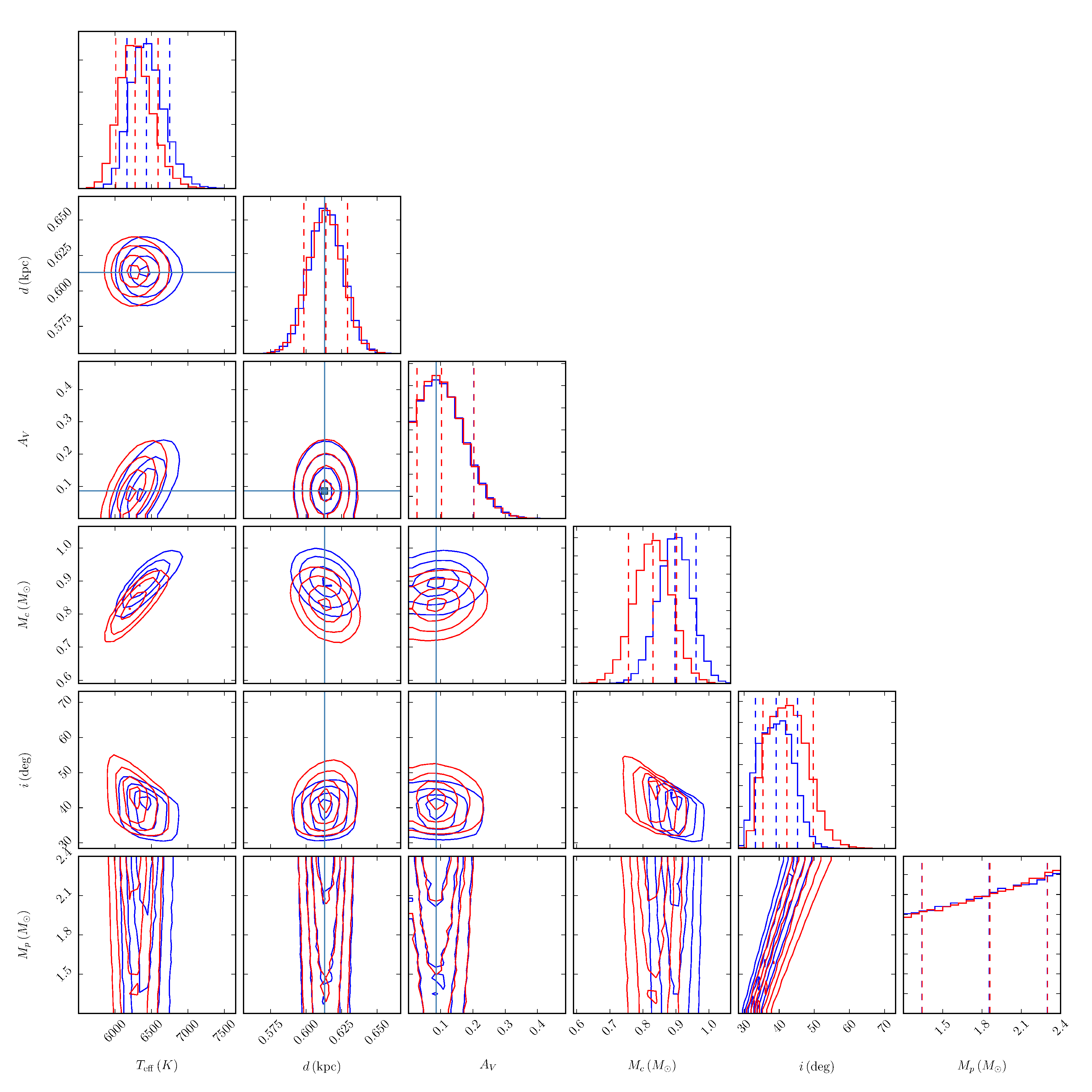}
	\caption{\label{fig:phot_fit2145}Joint two-dimentional posterior probability distribution
    functions and marginalized one-dimensional posterior probability 
    distribution functions from the MCMC analysis on \psrtwo. The prior 
    distributions are as described in Sec.~\ref{sec:discussion}. The 
    blue curves are for hydrogen-atmosphere (DA) WDs, while the red 
    curves are for helium-atmosphere (DB) WDs. The vertical dashed lines 
    show the median, 10\%, and 90\% confidence limits. The horizontal/vertical 
    solid lines show the nominal values of $d$ and $A_V$ from Table~\ref{tab:fit} 
    and \ref{tab:photometry}, respectively.}
	\end{center}
\end{figure*}

The best-fit parameter values are listed in Table~\ref{tab:phot_fits}, and the posterior 
probability distributions are shown in Figures~\ref{fig:phot_fit1022} and \ref{fig:phot_fit2145} 
for \psrone\ and \psrtwo, respectively.  Results using the DA models are in blue, 
while the DB models are in red.  Overall, the DB models have a slightly lower 
\Teff\ than the DA models, which requires a slightly larger radius (i.e., smaller \Mc) 
to match the \hst\ flux for a given distance. Although these plots were 
generated using models for a 0.8\,$M_\odot$ WD, using a 
0.7\,$M_\odot$ or 0.9\,$M_\odot$ WD changed the results by $\ll 1\,\sigma$.
Our WD temperatures are higher than those reported by \citet{lfc96}, but are 
consistent with the revised values from \citet{1998MNRAS.294..569H}.

\begin{deluxetable}{l c c}
  \tablewidth{0pt}
\tablecaption{Fit Results for WD and Pulsar Companions\label{tab:phot_fits}}
\tablehead{
  \colhead{Parameter} & \colhead{\psrone} & \colhead{\psrtwo}}
\startdata
\sidehead{DA (hydrogen atmosphere) WD}
\Teff\ (K) & $8128_{-346}^{+359}$ & $6441_{-209}^{+245}$\\
\Rc\ ($10^{-2}R_\odot$)& $0.89_{-0.05}^{+0.06}$ & $0.91_{-0.06}^{+0.06}$ \\
\Mc\ ($M_\odot$)& $0.92_{-0.05}^{+0.05}$ & $0.90_{-0.05}^{+0.05}$\\
$\tau_{\rm cool}$ (Gyr) & $2.7_{-0.2}^{+0.1}$ & $4.4_{-0.2}^{+0.2}$\\
$i$ (deg) & $66_{-10}^{+9}$ & $39_{-5}^{+5}$\tablenotemark{a} \\
\Mp\ ($M_\odot$) &$1.7_{-0.3}^{+0.3}$ & $1.8_{-0.4}^{+0.4}$ \\
$\chi^2$ & 0.5 & 0.9\\
\hline
\sidehead{DB (helium atmosphere) WD}
\Teff\ (K) & $7969_{-357}^{+404}$ & $6283_{-211}^{+240}$\\
\Rc\ ($10^{-2}R_\odot$)& $0.92_{-0.06}^{+0.06}$ & $0.96_{-0.06}^{+0.06}$ \\
\Mc\ ($M_\odot$)& $0.87_{-0.06}^{+0.06}$ & $0.83_{-0.06}^{+0.06}$\\
$\tau_{\rm cool}$ (Gyr) & $2.5_{-0.2}^{+0.1}$ & $4.4_{-0.2}^{+0.2}$\\
$i$ (deg) & $69_{-9}^{+8}$ & $42_{-5}^{+6}$ \\
\Mp\ ($M_\odot$) &$1.6_{-0.2}^{+0.3}$ & $1.8_{-0.4}^{+0.4}$ \\
$\chi^2$ & 0.2 & 0.5\\
\enddata
\tablecomments{Values are the median and 68\% confidence ranges from
  the marginalized posterior distributions.}
\end{deluxetable}

The very low $\chi^2$ values are a result of having more degrees of 
freedom than data points.  For instance, \Teff\ and $A_V$ are highly degenerate, 
as are \Teff\ and \Mc\ (or \Rc).  As long as our source photometry lies
in the color-magnitude region spanned by the models we will be able to
find an acceptable fit with only two photometric measurements, as is the
case for \psrone: the color determines \Teff\ (for a given $A_V$), and
the magnitude determines \Mc\ (for a given $d$).  With three
measurements, like we have for \psrtwo, it can be harder to find an acceptable fit, 
but the system is still over-determined and our models fit the data well.

\subsubsection{System inclinations and pulsar masses}
We note that in Figures~\ref{fig:phot_fit1022} and \ref{fig:phot_fit2145}, the posteriors for \Mp\ 
extend to high (and unphysical) masses.  We could have
eliminated this with an upper limit to the mass prior, but given that
we are looking to constrain possibly large masses we wished to avoid imposing 
a somewhat arbitrary prior.  Usually, with a guess
for \Mp\ there is a lower limit for \Mc\ \citep[e.g.,][]{lk12} which
occurs for $i=90\degr$.
However, in this case with a constraint on \Mc\ there is instead an upper
limit for \Mp\ which occurs for $i=90\degr$:
\[
M_{{\rm psr,max}} = \Mc \left( \sqrt{ \frac{\Mc T_\odot P_b^2}{4\pi^2
    x^3}}-1\right).
\]
Therefore, because of the largely uninformative prior we have on $i$, there
is considerable probability of a large inclination and hence a large
pulsar mass.  For each pulsar, we can impose only a weak upper limit on
$i$, using the combination of the VLBI proper motion with a timing 
measurement of $\dot{x}$.  For \psrone, \citet{reardon16a} and \citet{desvignes16a}
provide $\dot{x}$ results which differ by $>$3$\sigma$, and result in limits
of $i<80$\degrees\ and $i<70$\degrees\ (3$\sigma$ confidence in each case) 
respectively.  Since we have no reason to favour one timing result over the other, we
conservatively exclude only $i>80$\degrees. For \psrtwo, all recently published
$\dot{x}$ results are consistent, and exclude $i>74$\degrees.  We note that this 
is substantially different to the constraint of $i<61$\degrees\ previously calculated 
for \psrtwo\ by \citep{2004AA...426..631L} using older, lower-quality timing data.

In principle, we could have used the results of the photometric fitting for inclination to tighten the constraints on the VLBI fit for reflex motion, or we could have used the inclination fit from the VLBI reflex motion as a prior for the photometric fit.  We chose to keep these two fitting processes separate for simplicity and robustness, and examine the inclination constraints separately.  For both pulsars, fitting the VLBI reflex motion gives only weak constraints on $i$, which limits the utility of this comparison.  In each case the best-fit VLBI inclination favors a more face-on orientation, which if imposed as a prior on the photometric fit would lead to a less massive pulsar and a (marginally) hotter and more massive companion, with shifts of up to $\sim$1$\sigma$ in \Mp\ and less in \Mc\ and \Teff.  We note also that \psrone\ has a marginal Shapiro delay measurement of $\mathrm{sin}\ i = 0.69 \pm 0.18$ \citep{reardon16a}, which also favours a somewhat more face-on geometry than the best-fit value from the photometric modeling. Alternatively, if the results from the photometric fitting had been used to tighten the allowable range of $i$ in the VLBI fit of reflex motion, then for both pulsars the uncertainties in parallax, proper motion, and $\Omega$ would have been reduced by up to 30\%, with shifts in the best-fit parallax and proper motion of $<1\sigma$.

\subsubsection{The age and mass of the white dwarf companions}
\begin{figure}
  \includegraphics[width=0.5\textwidth]{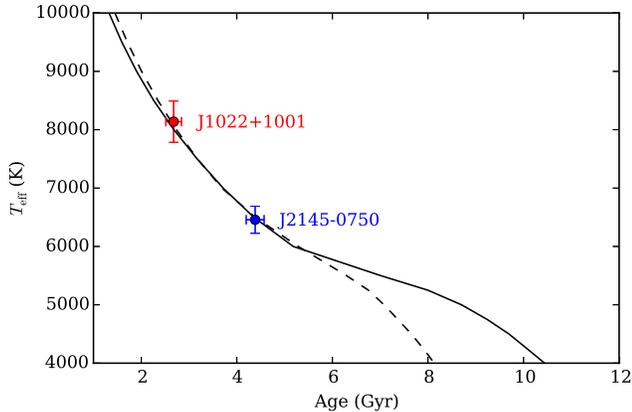}
  \caption{\label{fig:cool}White dwarf temperatures and ages for \psrone\ (red) and \psrtwo\ (blue) 
  assuming a hydrogen atmosphere.
   The lines indicate cooling curves are those for a
    $0.9\,M_\odot$ CO WD with a hydrogen (solid) or helium (dashed)
    atmosphere.
  }
\end{figure}

The temperatures and ages of \psrone\ and \psrtwo\ are plotted in Fig.~\ref{fig:cool}, 
along with cooling curves for DA and DB CO WDs with mass 
of $0.9\:M_\odot$. Both sources are relatively hot, and therefore relatively young, 
with ages of about 2 Gyr and 4 Gyr, respectively.  For both of these sources, the 
cooling ages are much younger than the pulsar's characteristic spin-down age, 
\[
\tau_c = -\frac{1}{n-1} \frac{P}{\dot{P}} \,,
\]
where $P$ is the spin period, $\dot{P}$ is the spin period derivative, and $n$ is the 
braking index.  The characteristic age assumes an initial spin period much shorter than the 
present value; if we assume that the pulsar spin-down is solely due to magnetic dipole 
braking ($n=3$) and that their true ages correspond to the white dwarf ages, 
then we find that the initial spin of these pulsars was very close to the 
present value, and they were only mildly recycled.  This is in line with evolutionary models,
which suggest that binary MSPs with massive WD companions are formed in intermediate-mass
X-ray binaries with a relatively short Roche-lobe overflow phase, where the limited 
amount of mass transfer leads to a typical spin period of tens of milliseconds \citep{tauris12a}.

\subsection{Comparing timing and VLBI astrometry}
\label{sec:vlbivstiming}

The VLBI astrometric results also allow an independent check of the astrometric parameters derived from pulsar timing.  Comparisons of the reference position are presently of limited value, since the error budget of the absolute position of our targets is dominated by uncertainties in the reference positions and frequency-dependent structure of our calibrator sources.  A dedicated campaign utilising multi-frequency observations and multiple primary calibrators could reduce this uncertainty to a few tenths of a milliarcsecond and enable precision tests of the alignment of the quasi-inertial reference frame used for VLBI observations \citep[e.g., the International Celestial Reference Frame;][]{ma09a} and the solar system barycentric frame used for pulsar timing; we intend to pursue this with future observations.  The absolute and time-invariant positional uncertainty of the in-beam calibrators does not, however, impact comparisons of proper motion and parallax, which we examine below.

Second order effects such as unmodeled time-varying structure of the calibrator sources, which would be transferred to the target pulsar and lead to a proper motion and/or parallax error, are typically small compared to the measurement error: an analysis of radio AGN comparable to our in-beam calibrators showed $\lesssim20\mu$as yr$^{-1}$ for 80\% of sources, although one source out of 61 exceeded 100 $\mu$as yr$^{-1}$ \citep{moor11a}.  For the pulsar timing, imperfect modeling of effects such as time-dependent dispersion measure (DM) variations caused by the changing line-of-sight through the turbulent interstellar medium can likewise lead to a systematic bias in estimates of proper motion and (especially) parallax.  Of particular concern are pulsars at low ecliptic latitude; here, the line of sight regularly passes close to the Sun, which will introduce annually-modulated errors if propagation effects in the solar wind are not correctly modeled and removed \citep{lam16a}.  We note that VLBI astrometry is not expected to be biased by the target's ecliptic latitude, since observations are always scheduled near the time of maximum parallax signature, when the angular separation between the target and Sun is $\sim$90\degrees.

Tables~\ref{tab:j1022pta} and~\ref{tab:j2145pta} show a comparison between the VLBI results presented here and the most recent pulsar timing results presented by PTAs.  The EPTA results are taken from \citet{desvignes16a}, the NANOGrav results from \citet{matthews16a}, and the PPTA results from \citet{reardon16a}. The higher precision on individual VLBI measurements means that despite the shorter time baseline, the errors on the VLBI proper motion are comparable to or better than the pulsar timing results, while the VLBI parallax measurements have an order of magnitude higher precision than the timing parallax measurements. 

\psrone\ is observed by the EPTA and PPTA, but not by NANOGrav.  Its location in the ecliptic plane (ecliptic latitude 0.06\degrees) complicates the measurement of positions via pulsar timing: the highly elongated error ellipse for position in ecliptic coordinates means that the position and proper motion errors are are highly covariant when quoted in equatorial coordinates.  This is a plausible cause of the large errors in the timing astrometric parameters highlighted in Table~\ref{tab:j1022pta}: for the EPTA, the derived parallax is in error by 3.5$\sigma$ compared to the much more precise VLBI value, while the PPTA measurement of proper motion in right ascension differs by 70$\sigma$ from the VLBI value.  However, it appears that the PPTA proper motion uncertainty in \citet{reardon16a} is underestimated due to the covariance with the poorly contrained proper motion in declination, meaning the significance of the discrepancy is greatly overstated (D. Reardon, priv. comm.)  We further note that excess timing noise dependent on the observing system was noted in the combined International Pulsar Timing Array analysis of \citet{verbiest16a}, which may also play a role in the timing astrometry errors.  Fixing the timing proper motion and parallax to the VLBI values will improve the timing model of this pulsar, and should thereby improve the contribution of this pulsar to the respective PTA sensitivities to gravitational waves \citep{madison13a}.

\psrtwo\ is observed by all three PTAs.  The agreement in fitted parallax is reasonable, with all 3 PTA values differing from the more accurate VLBI value by $\lesssim1.5\sigma$.  Regarding proper motion, the EPTA and PPTA results also agree at the $\sim$2$\sigma$ level or better; however the NANOgrav results differ from the more precise VLBI values by 3--5 $\sigma$.  NANOGrav has a shorter timing baseline (9 years) than the EPTA (17.5 years) and PPTA (17 years), which when combined with DM variation modeling may lead to larger systematic proper motion errors.  Although not as close to the ecliptic plane as \psrone, \psrtwo\ also has a relatively low ecliptic latitude (5.3\degrees) leading to similar albeit less severe covariance issues when reporting proper motion in equatorial coordinates, and long term timing noise was also noted by \citet{verbiest16a}.
We note that other NANOGrav pulsar has existing VLBI astrometry: PSR J1713+0747, observed by \citet{chatterjee09a}.  For this system, which is at a much higher ecliptic latitude of 30.7\degrees, consistency between the VLBI and timing proper motions is seen. 

\begin{deluxetable}{lccc}
\tabletypesize{\scriptsize}
\tablecaption{\label{tab:j1022pta}Timing vs.\ VLBI astrometry for \psrone\tablenotemark{A}.}
\tablehead{ \colhead{Origin} & \colhead{Proper motion} & \colhead{Proper motion)} & \colhead{Parallax} \\
 & \colhead{(R.A., mas/yr)} & \colhead{Decl. mas/yr)} & \colhead{(mas)} }
\startdata
This work 		& $-$14.86(4)	& 5.59(3)	& 1.43$^{+0.02}_{-0.03}$		\\
EPTA		& $-$18.2(64)	& 3(16)	& 0.72(20) 	\\
PPTA		& $-$17.09(3)	& --		& 1.1(3)		
\enddata
\tablenotetext{A}{Timing proper motion in R.A.\ and Dec.\ are highly correlated due to the location in the ecliptic plane.}
\end{deluxetable}

\begin{deluxetable}{lccc}
\tabletypesize{\scriptsize}
\tablecaption{\label{tab:j2145pta}Timing vs.\ VLBI astrometry for \psrtwo.}
\tablehead{ \colhead{Origin} & \colhead{Proper motion} & \colhead{Proper motion)} & \colhead{Parallax} \\
 & \colhead{(R.A., mas/yr)} & \colhead{Decl. mas/yr)} & \colhead{(mas)} }
\startdata
This work 		& $-$9.46(5)	& $-$9.08(6)	& 1.63(4)		\\
EPTA		& $-$9.58(4)	& $-$8.86(4)	& 1.53(11) 	\\
NANOGrav	& $-$10.1(1)	& $-$7.5(4)	& 1.3(2)		\\
PPTA		& $-$9.59(8)	& $-$8.9(3)	& 1.84(17)		
\enddata
\end{deluxetable}

\begin{figure*}
\begin{center}
\begin{tabular}{cc}
\includegraphics[width=0.49\textwidth]{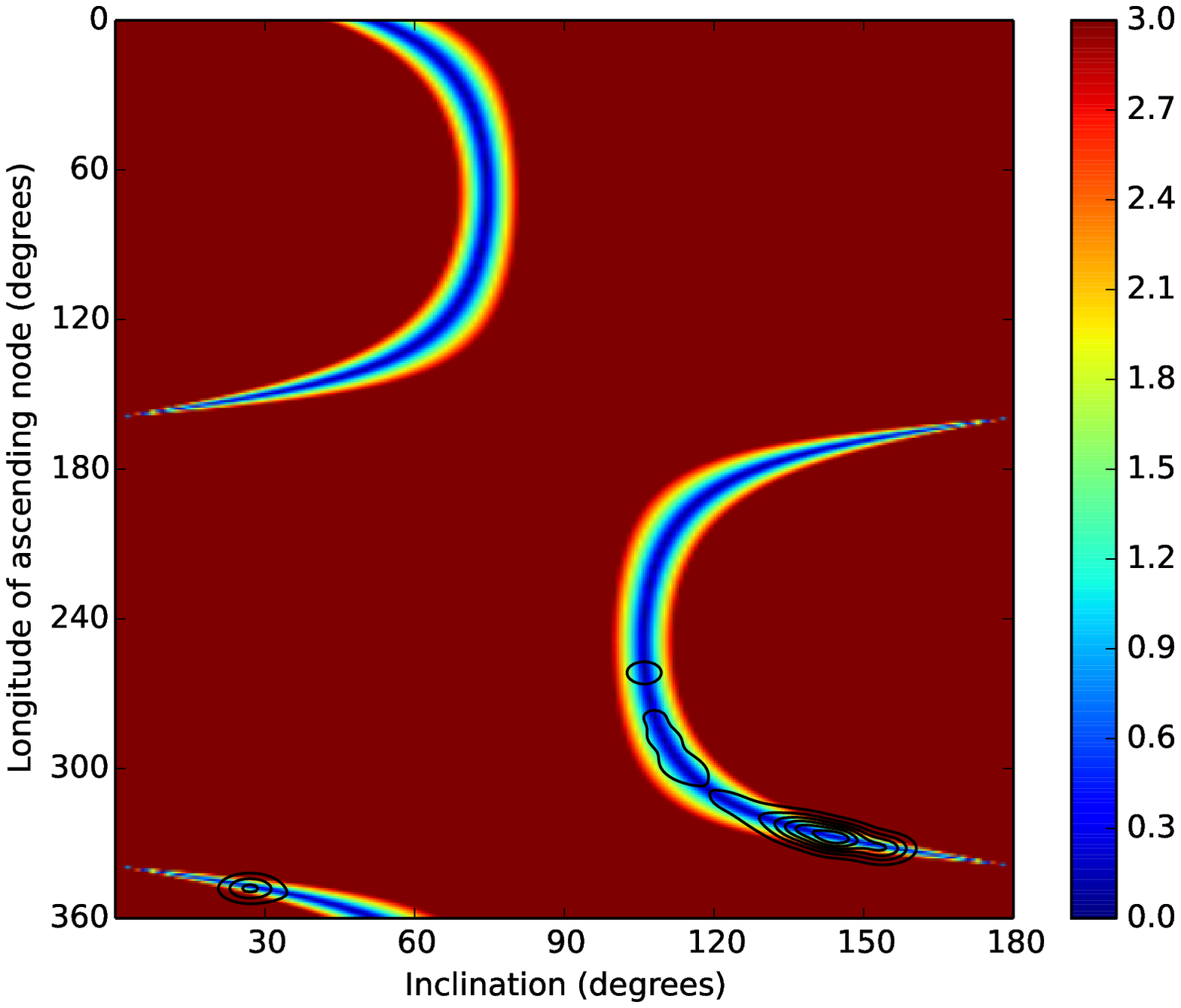} & \includegraphics[width=0.49\textwidth]{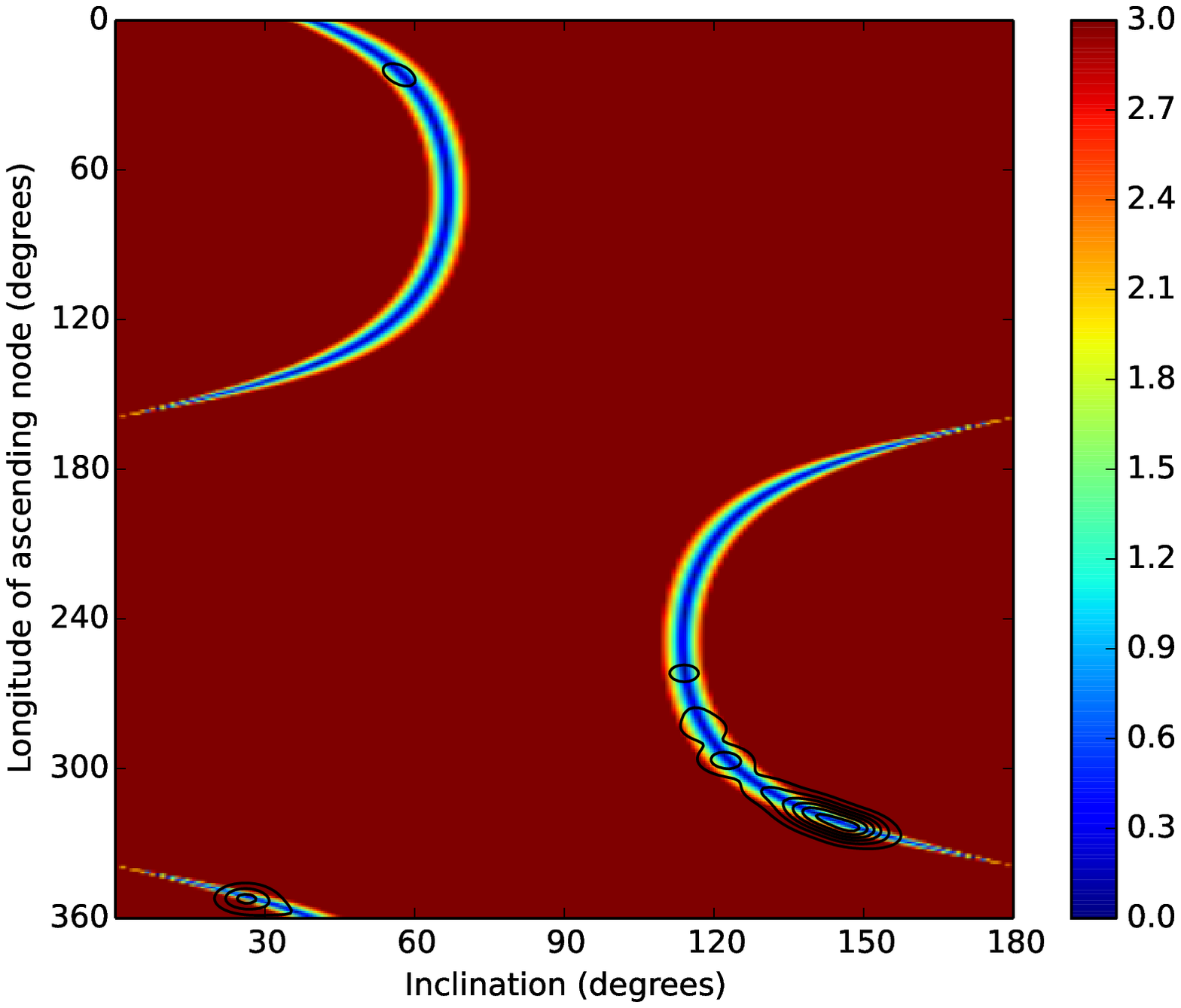}  \\
\includegraphics[width=0.49\textwidth]{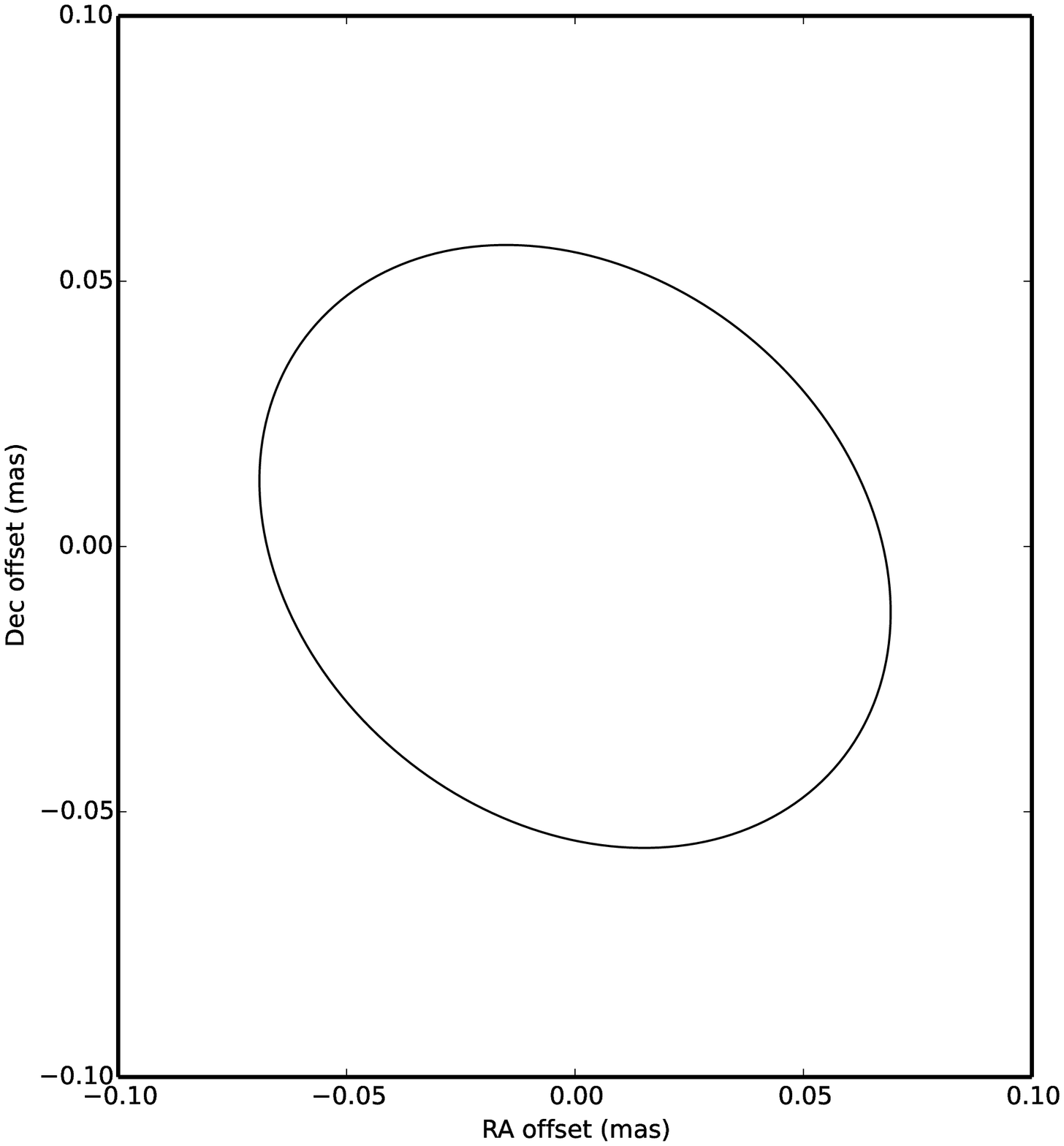}  & \includegraphics[width=0.49\textwidth]{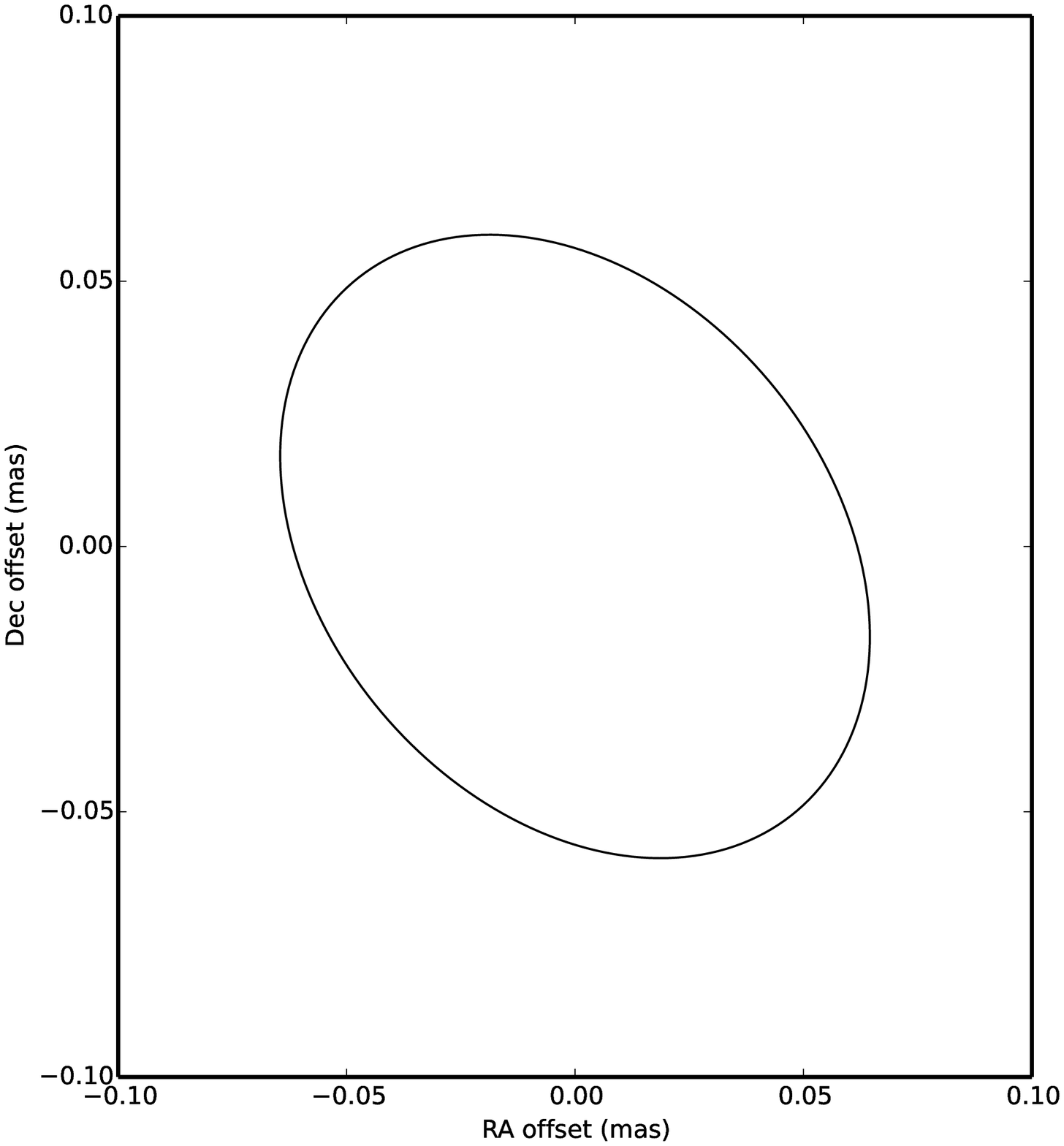} 
\end{tabular}
\end{center}
\caption{\label{fig:xdoteffect}The orbital parameters and motion of \psrone, using $\dot{x}$ constraints from the PPTA \citep[left column]{reardon16a}, and EPTA \citep[right column]{desvignes16a}.  The top row shows the allowed region of $(\Omega, i)$ parameter space, with the color scale showing deviation from the timing $\dot{x}$ value in standard deviations, and the contours showing the density of fitted points from the VLBI bootstrap.  Despite the large change in $\dot{x}$ between the columns, the fitted values of $i$ and $\Omega$ from the VLBI reflex motion do not change greatly: using the EPTA $\dot{x}$ changes $\Omega$ from 336\degrees\ to 324\degrees, and $i$ from 138\degrees\ to 149\degrees. The bottom row shows the reflex motion on the sky resulting from the most likely combination of $\Omega$ and $i$ in each case, showing that the reflex motion is not greatly affected by the choice of $\dot{x}$ constraint.}
\end{figure*}

The final area of discrepancy between PTA results is the published $\dot{x}$ values for \psrone.  Using the EPTA value of $(1.79 \pm 0.12) \times 10^{-14}$ \citep{desvignes16a} rather than the PPTA value of $(1.15 \pm 0.16) \times 10^{-14}$ \citep{reardon16a} leads to changes of $\ll1\sigma$ in the VLBI astrometric parameters.  This can be understood by considering that the altered value of $\dot{x}$ can be fit by quite small changes in $i$ and $\Omega$, which lead to only a very small perturbation of the (already small) reflex motion on the sky, as illustrated in Figure~\ref{fig:xdoteffect}.  Discerning which of the two values of $\dot{x}$ is correct from the reflex motion alone would require VLBI astrometry with an order of magnitude higher precision, capable of fitting $\Omega$ to a precision of a few degrees without using the $\dot{x}$ constraints.  This is likely impossible with current instrumentation, but may be possible in the future with the incorporation of the Square Kilometre Array into VLBI networks \citep{paragi15a}.

\section{Conclusions}
Using VLBI astrometry, we have measured the distance and transverse velocity of the binary millisecond pulsars \psrone\ and \psrtwo\ with a precision of $\sim$2\%.  Our astrometric results show that even state-of-the-art pulsar timing can significantly underestimate the uncertainty of timing parallax and proper motion measurements in unfavourable cases (such as at low ecliptic latitude), and reiterate that distance estimates based on the pulsar dispersion measure and Galactic electron density distribution models should be treated with due caution for individual systems.  We use the precise VLBI distances along with a revised and improved analysis of \hst\ photometry to calculate the mass, radius, and effective temperature of the white dwarf companions to \psrone\ and \psrtwo\ with a precision of around 10\%.  Together, these give independent age and velocity constraints that can be used to evaluate and improve MSP formation and evolution models. The current photometric data cannot constrain the companion atmosphere composition, and helium atmosphere models give a companion mass that differs (albeit by $<1\,\sigma$) from hydrogen atmosphere models.  Future optical spectroscopy should be able to determine which atmosphere model is correct, as well as measure the surface gravity.  This would give another constraint on \Rc\ and \Mc\ to further tighten the range of possible solutions for these systems.

\acknowledgements  ATD was supported by an NWO Veni Fellowship. DLK and SJV receive support from the NANOGrav project through National Science Foundation (NSF) PIRE program award number 0968296 and NSF Physics Frontiers Center award number 1430284.  Part of this research was carried out at the Jet Propulsion Laboratory, California Institute of Technology, under a contract with the National Aeronautics and Space Administration. The authors thank David Nice and Pierre Bergeron for useful discussions. The National Radio Astronomy Observatory is a facility of the National Science Foundation operated under cooperative agreement by Associated Universities, Inc.  Pulsar research at the Jodrell Bank Centre for Astrophysics and the observations using the Lovell Telescope are supported by a consolidated grant from the STFC in the UK.  Funding for SDSS-III has been provided by the Alfred P. Sloan Foundation, the Participating Institutions, the National Science Foundation, and the U.S. Department of Energy Office of Science. The SDSS-III web site is http://www.sdss3.org/.

\bibliographystyle{apj}
\bibliography{deller_thesis,wd_hst}

\end{document}